\input harvmac
\input epsf
\noblackbox
%%%%%%%%%%%%%%%%%%%%%%%%%%%%%%%%%%%%%%%%%%%%%%%%%%%%%%%%%%%%%%%%%%%%%%%%%
\def\npb#1#2#3{{\it Nucl.\ Phys.} {\bf B#1} (#2) #3}
\def\plb#1#2#3{{\it Phys.\ Lett.} {\bf B#1} (#2) #3}
\def\prl#1#2#3{{\it Phys.\ Rev.\ Lett.} {\bf #1} (#2) #3}
\def\prd#1#2#3{{\it Phys.\ Rev.} {\bf D#1} (#2) #3}
\def\rmp#1#2#3{{\it Rev.\ Mod.\ Phys.} {\bf #1} (#2) #3}
\def\mpla#1#2#3{{\it Mod.\ Phys.\ Lett.} {\bf A#1} (#2) #3}
\def\ijmp#1#2#3{{\it Int.\ J. Mod.\ Phys.} {\bf A#1} (#2) #3}
\def\cmp#1#2#3{{\it Commun.\ Math.\ Phys.} {\bf #1} (#2) #3}

\def\jmp#1#2#3{{\it J. Math.\ Phys.} {\bf #1} (#2) #3}

\def\jhep#1#2#3{{\it JHEP\/} {\bf #1} (#2) #3}
\def\atmp#1#2#3{{\it Adv.\ Theor.\ Math.\ Phys.} {\bf #1} (#2) #3}
%%%%%%%%%%%%%%%%%%%%%%%%%%%%%%%%%%%%%%%%%%%%%%%%%%%%%%%%%%%%%%%%%%%%%%%%%
% some stuff needed for figures:
%%%%%%%%%%%%%%%%%%%%%%%%%%%%%%%%%%%%%%%%%%%%%%%%%%%%%%%%%%%%%%%%%%%%%%%%%
\newcount\figno
\figno=0
\def\fig#1#2#3{
\par\begingroup\parindent=0pt\leftskip=1cm\rightskip=1cm\parindent=0pt
\baselineskip=11pt
\global\advance\figno by 1
\midinsert
\epsfxsize=#3
\centerline{\epsfbox{#2}}
\vskip 12pt
{\bf Fig.\ \the\figno: } #1\par
\endinsert\endgroup\par
}
\def\figlabel#1{\xdef#1{\the\figno}}
\def\encadremath#1{\vbox{\hrule\hbox{\vrule\kern8pt\vbox{\kern8pt
\hbox{$\displaystyle #1$}\kern8pt}
\kern8pt\vrule}\hrule}}
%%%%%%%%%%%%%%%%%%%%%%%%%%%%%%%%%%%%%%%%%%%%%%%%%%%%%%%%%%%%%%%%%%%%%%%%%
\def\bberkeley{\centerline{\it Berkeley Center for Theoretical Physics and 
Department of Physics}
\centerline{\it University of California, Berkeley, CA 94720-7300}
\centerline{\it and}
\centerline{\it Theoretical Physics Group, Lawrence Berkeley National 
Laboratory}
\centerline{\it Berkeley, CA 94720-8162, USA}}
%%%%%%%%%%%%%%%%%%%%%%%%%%%%%%%%%%%%%%%%%%%%%%%%%%%%%%%%%%%%%%%%%%%%%%%%%

\def\frac#1#2{{#1 \over #2}}

\def\p{\partial}
\def\semi{\subset\kern-1em\times\;}

\def\CH{{\cal H}}

\def\CO{{\cal O}}                   
\def\CR{{\cal R}}                   
                   
                   \def\CZ{{\cal Z}}
\def\C{{\bf C}}                     
\def\R{{\bf R}}          %          \def\S{{\bf S}}
                     \def\Z{{\bf Z}}
\def\Re{{\,{\rm Re}\,}}
\def\Im{{\,{\rm Im}\,}}
\def\Li{{\,\rm Li}}

\def\ie{{\it i.e.}}
\def\eg{{\it e.g.}}

\def\mudual{\widetilde{\mu}}
\def\hepth#1{[arXiv:hep-th/{#1}]}
%
%%%%%%%%%%%%%%%%%%%%%%%%%%%%%%%%%%%%%%%%%%%%%%%%%%%%%%%%%%%%%%%%%%%%%%%%%%%
\Title{\vbox{\baselineskip12pt
\hbox{hep-th/0512325}
\hbox{\ }
\hbox{\ }}}
{\vbox{\centerline{Thermodynamics of Noncritical M-Theory}
\bigskip
\centerline{and the Topological A-Model}}}
%%%%%%%%%%%%%%%%%%%%%%%%%%%%%%%%%%%%%%%%%%%%%%%%%%%%%%%%%%%%%%%%%%%%%%%%%%%
\font\authfont=cmr12
\centerline{\authfont Petr Ho\v rava and Cynthia A. Keeler}
\medskip\bigskip
\baselineskip14pt
\bberkeley
\medskip\bigskip\medskip
\centerline{\bf Abstract}
\bigskip
\noindent
In hep-th/0508024, noncritical M-theory for two-dimensional Type 0A and 
0B strings was defined in terms of a double-scaled theory of nonrelativistic 
fermions in $2+1$ dimensions.  Here we study this noncritical M-theory at 
finite temperature.  We derive the exact expression for the free energy of 
its vacuum solution, as a function of a coupling constant $g_M$ and the radius 
$R$ of the thermal circle.  We show that at high temperature, the theory is 
effectively described by another M-theory solution, whose effective 
loop-counting coupling scales in a novel way characteristic of M-theory, as 
$T^3$.  Our calculations further suggest that noncritical M-theory is dual to 
the closed string theory of the topological A-model on a Calabi-Yau, 
with the radius $R$ of the Euclidean time circle in M-theory playing the role 
of the string coupling constant of the A-model.  In this correspondence, 
T-duality on the Euclidean time circle of noncritical M-theory implies 
an S-duality for the topological A-model.  
\Date{December 2005}
%%%%%%%%%%%%%%%%%%%%%%%%%%%%%%%%%%%%%%%%%%%%%%%%%%%%%%%%%%%%%%%%%%%%%%%%%%%
\nref\ncm{P. Ho\v rava and C.A. Keeler, ``Noncritical M-Theory in $2+1$ 
Dimensions as a Nonrelativistic Fermi Liquid'' \hepth{0508024}.}
\nref\nakarev{Y. Nakayama, ``Liouville Field Theory -- A Decade after the 
Revolution'' \hepth{0402009}.}
\nref\emilrev{E.J. Martinec, ``The Annular Report on Non-Critical String 
Theory'' \hepth{0305148}, ``Matrix Models and 2D String Theory'' 
\hepth{0410136}.}
\nref\alexrev{S. Alexandrov, ``Matrix Quantum Mechanics and Two-dimensional 
String Theory in Non-trivial Backgrounds'' \hepth{0311273}.}
\nref\sumit{S.R. Das, ``Non-trivial 2d Space-Times from Matrices'' 
\hepth{0503002}.}
\nref\polrev{J. Polchinski, ``What is String Theory?'' \hepth{9411028}.}
\nref\igrev{I. Klebanov, ``String Theory in Two Dimensions'' \hepth{9108019}.}
\nref\gmrev{P. Ginsparg and G. Moore, ``Lectures on 2D Gravity and 2D String 
Theory'' \hepth{9304011}.}
\nref\newhat{M.R.~Douglas, I.R.~Klebanov, D.~Kutasov, J.~Maldacena and 
E.~Martinec, ``A New Hat for the $c=1$ Matrix Model'' \hepth{0307195}.}
\nref\tato{T. Takayanagi and N. Toumbas, ``A Matrix Model Dual of Type 0B 
String Theory in Two Dimensions,'' \jhep{0307}{2003}{064} \hepth{0307083}.}
\nref\grossm{D.J. Gross and P. Mende, ``The High-Energy Behavior of String 
Scattering Amplitudes,'' \plb{197}{1987}{129}; 
``String Theory Beyond the Planck Scale,'' \npb{303}{1988}{407}.}
\nref\amcave{D. Amati, M. Ciafaloni and G. Veneziano, ``Superstring Collisions 
at Planckian Energies,'' \plb{197}{1987}{81}; ``Classical and Quantum Gravity 
Effects from Planckian Energy Superstring Collisions,'' \ijmp{3}{1988}{1615}.}
\nref\various{D.J. Gross, ``High-Energy Symmetries of String Theory,'' 
\prl{60}{1988}{1229}\hfill\break
E. Witten, ``Space-Time and Topological Orbifolds,'' \prl{61}{1988}{670}
\hfill\break
D. Amati, M. Ciafaloni and G. Veneziano, ``Can Spacetime be Probed below the 
String Size?'' \plb{216}{1989}{41}.}
\nref\aw{J.J. Atick and E. Witten, ``The Hagedorn Transition and the 
Number of Degrees of Freedom of String Theory,'' \npb{310}{1988}{334}.}
\nref\atat{G.G. Athanasiu and J.J. Atick, ``Remarks on Thermodynamics of 
Strings,'' in: {\it Strings '88: Proceedings}, eds: S.J. Gates et al.\ 
(World Scientific, 1989).}
\nref\appr{P. Ho\v rava and C.A. Keeler, ``Strings on $AdS_2$ and the 
High-Energy Limit of Noncritical M-Theory,'' to appear.}
\nref\alvaos{E. Alvarez and M.A.R. Osorio, ``Superstrings at Finite 
Temperature,'' \prd{36}{1987}{1175}.}
\nref\mcguigan{M. McGuigan, ``Finite-Temperature String Theory and Twisted 
Tori,'' \prd{38}{1988}{552}.}
\nref\bowick{M.J. Bowick and L.C.R. Wijewardhana, ``Superstrings at High 
Temperature,'' \prl{54}{1985}{2485}\hfill\break
M.J. Bowick and S.B. Giddings, ``High-Temperature Strings,'' 
\npb{325}{1989}{631}.}
\nref\tan{N. Deo, S. Jain and C.-I. Tan, ``Strings at High-Energy Densities 
and Complex Temperature,'' \plb{220}{1989}{125}; ``The Ideal Gas of Strings,'' 
in: {\it Modern Quantum Field Theory: Proceedings}, eds: S. Das et al.\ 
(World Scientific, 1991).} 
\nref\physica{{\it Thermal Field Theories and Their Applications}, eds: 
K.L.~Kowalski et al.,~{\it Physica\/} {\bf A158} (1989) No.~1.}
\nref\russo{J.G. Russo, ``Free Energy and Critical Temperature in Eleven 
Dimensions'' \hepth{0101132}.}
\nref\colum{R. Easther, B.R. Greene, M.G. Jackson and D. Kabat, ``Brane Gases 
in the Early Universe: Thermodynamics and Cosmology'' \hepth{0307233}.}
\nref\oren{O. Bergman and M. Gaberdiel, ``Dualities of Type 0 Strings,'' 
\jhep{9907}{1999}{022} \hepth{9906055}.}
\nref\topomth{R. Dijkgraaf, S. Gukov, A. Neitzke and C. Vafa, ``Topological 
M-Theory as Unification of Form Theories of Gravity'' \hepth{0411073}.}
\nref\topozth{N. Nekrasov, ``Z-Theory'' \hepth{0412021}.}
\nref\malsei{J. Maldacena and N. Seiberg, ``Flux-vacua in Two Dimensional 
String Theory'' \hepth{0506141}.}
\nref\gova{R. Gopakumar and C. Vafa, ``On the Gauge Theory/Geometry 
Correspondence,'' \atmp{3}{1999}{1415} \hepth{9811131}.} 
\nref\govam{R. Gopakumar and C. Vafa, ``M-Theory and Topological 
Strings -- I and II'' \hepth{9809187} and \hepth{9812127}.}
\nref\foam{A. Okounkov, N. Reshetikhin and C. Vafa, ``Quantum Calabi-Yau and 
Classical Crystals'' \hepth{0309208}\hfill\break
A. Iqbal, N. Nekrasov, A. Okounkov and C. Vafa, ``Quantum Foam and Topological 
Strings'' \hepth{0312022}.}
\nref\grosskl{D.J. Gross and I.R. Klebanov, ``One-Dimensional String Theory 
on a Circle,'' \npb{344}{1990}{475}.}
\nref\hetm{P. Ho\v rava and E. Witten, ``Heterotic and Type I String Dynamics 
from Eleven Dimensions,'' \npb{460}{1996}{5-6} \hepth{9510209}; ``Eleven 
Dimensional Supergravity on a Manifold with Boundary,'' \npb{475}{1996}{94} 
\hepth{9603142}.}
\nref\shenker{S. Shenker, ``The Strength of Nonperturbative Effects in String 
Theory,'' in: {\it Random Surfaces and Quantum Gravity}, Carg\`ese 1990 
Proceedings.}
\nref\grossw{D.J. Gross and J. Walcher, ``Non-perturbative RR Potentials 
in the $\hat c=1$ Matrix Model'' \hepth{0312021}.}
\nref\gtt{S. Gukov, T. Takayanagi and N. Toumbas, ``Flux Backgrounds in 
2D String Theory'' \hepth{0312208}.}
\nref\ulf{U.H. Danielsson, M.E. Olsson and M. Vonk, ``Matrix Models, 4D Black 
Holes and Topological Strings on Non-compact Calabi-Yau Manifolds'' 
\hepth{0410141}.}
\nref\toporevmm{M. Mari\~ no, ``Chern-Simons Theory and Topological 
Strings,'' \rmp{77}{2005}{675} \hepth{0406005}; ``Enumerative Geometry and 
Knot Invariants'' \hepth{0210145}; ``Les Houches Lectures on Matrix Models 
and Topological Strings'' \hepth{0410165}.}
\nref\toporevnv{A. Neitzke and C. Vafa, ``Topological Strings and Their 
Physical Applications'' \hepth{0410178}.}
\nref\penner{J. Distler and C. Vafa, ``A Critical Matrix Model at $c=1$,'' 
\mpla{6}{1991}{259}.}
\nref\oovafa{H. Ooguri and C. Vafa, ``Worldsheet Derivation of a Large $N$ 
Duality'' \hepth{0205297}.}
\nref\csmm{M. Mari\~ no,  ``Chern-Simons Theory, Matrix Integrals, and 
Perturbative Three-Manifold Invariants,'' \cmp{253}{2004}{25} 
\hepth{0207096}\hfill\break
M. Aganagic, A. Klemm, M. Mari\~ no and C. Vafa, ``Matrix Model as a Mirror 
of Chern-Simons Theory,'' \jhep{0402}{2004}{010} \hepth{0211098}.}
\nref\osv{H. Ooguri, A. Strominger and C.Vafa, ``Black Hole Attractors and 
the Topological String,'' \prd{70}{2004}{106007} \hepth{0405146}.}
\nref\vafaq{C. Vafa, ``Two Dimensional Yang-Mills, Black Holes and 
Topological Strings'' \hepth{0406058}.}
\nref\aosv{M. Aganagic, H. Ooguri, N. Saulina and C. Vafa, ``Black Holes, 
$q$-Deformed 2d Yang-Mills, and Non-perturbative Topological Strings'' 
\hepth{0411280}.}
\nref\grosst{D.J. Gross, ``Two Dimensional QCD as a String Theory,'' 
\npb{400}{1993}{161} \hepth{9212149};\hfill\break
D.J. Gross and W. Taylor, ``Two Dimensional QCD is a String 
Theory,'' \npb{400}{1993}{181} \hepth{9301068}; ``Twists and Wilson Loops 
in the String Theory of Two-Dimensional QCD,'' \npb{403}{1993}{395} 
\hepth{9303046}.}
\nref\rigid{P. Ho\v rava, ``Topological Strings and QCD in Two Dimensions,'' 
in: {\it Quantum Field Theory and String Theory}, 
Carg\`ese proceedings, 1993 \hepth{9311156}\hfill\break 
``Topological Rigid String Theory and Two-Dimensional QCD,'' 
\npb{463}{1996}{238} \hepth{9507060}; ``On QCD String Theory and AdS 
Dynamics,'' \jhep{9901}{1999}{016} \hepth{9811028}.}
\nref\cmr{S. Cordes, G. Moore and S. Ramgoolam, ``Large $N$ 2D Yang-Mills 
Theory and Topological String Theory,'' \cmp{185}{1997}{543} 
\hepth{9402107}.}
\nref\doukaz{M.R. Douglas and V.A. Kazakov, ``Large $N$ Phase Transition 
in Continuum QCD${}_2$,'' \plb{319}{1993}{219} \hepth{9305047}.}
\nref\dkosv{X. Arsiwalla, R. Boels, M. Mari\~no and A. Sinkovics, 
``Phase Transitions in $q$-Deformed 2d Yang-Mills Theory and Topological 
Strings'' \hepth{0509002}\hfill\break
D. Jafferis and J. Marsano, ``A DK Phase Transition in 
$q$-Deformed Yang-Mills on $S^2$ and Topological Strings'' 
\hepth{0509004}\hfill\break
N. Caporaso, M. Cirafici, L. Griguolo, S. Pasquetti, D. Seminara and R.J. 
Szabo, ``Topological Strings and Large $N$ Phase Transitions I and II'' 
\hepth{0509041} and \hepth{0511043}.}
\nref\neitzvafa{A. Neitzke and C. Vafa, ``N=2 Strings and the Twistorial 
Calabi-Yau'' \hepth{0402128}.}
\nref\toposdual{N. Nekrasov, H. Ooguri and C. Vafa, ``S-Duality and 
Topological Strings,'' \jhep{0410}{2004}{009} \hepth{0403167}.}
\nref\skapustin{A. Kapustin, ``Gauge Theory, Topological Strings, and 
S-Duality,'' \jhep{0409}{2004}{034} \hepth{0404041}.}
\nref\okuda{T. Okuda, ``Derivation of Calabi-Yau Crystals from Chern-Simons 
Gauge Theory,'' \jhep{0503}{2005}{047} \hepth{0409270}.}
\nref\atmava{M. Atiyah, J. Maldacena and C. Vafa, ``An M-Theory Flop as 
a Large N Duality,'' \jmp{42}{2001}{3209} \hepth{0011256}.}
\nref\ghvafa{D. Ghoshal and C. Vafa, ``$c=1$ String and the Topological 
Theory of the Conifold,'' \npb{453}{1995}{121} \hepth{9506122}.}
\nref\mina{M. Aganagic, R. Dijkgraaf, A. Klemm, M. Mari\~no and C. Vafa, 
``Topological Strings and Integrable Hierarchies'' \hepth{0312085}.}
%%%%%%%%%%%%%%%%%%%%%%%%%%%%%%%%%%%%%%%%%%%%%%%%%%%%%%%%%%%%%%%%%%%%%%%%%%%
\newsec{Introduction and Summary}

Noncritical string theories in two spacetime dimensions (see \refs{\nakarev-
\gmrev} for reviews) provide a useful laboratory in which many features of 
full string theory can be studied in a simpler, exactly solvable setting.  One 
aspect of the full theory which continues to be shrouded in mystery is its 
M-theory regime, where fundamental strings are no longer the correct degrees 
of freedom.  In this regime, the dynamics of the theory can be described at 
low energies by an effective supergravity theory, with M2-branes and M5-branes 
as solitons.  When combined with anticipated dualities to various string 
vacua, this rather incomplete picture still leads to a wealth of useful 
information about M-theory.  However, a more direct understanding of the 
M-theory degrees of freedom, in which one would not have to resort to a 
nonperturbative duality, is clearly desirable.  

This problem appears to be difficult in full eleven-dimensional M-theory, 
but it is precisely the type of question that could be addressed first 
in the simplified setting of two-dimensional string theory.  Type 0A and 0B 
strings in two dimensions \refs{\newhat,\tato} seem particularly suitable for 
this purpose.  The spectrum of two-dimensional Type 0A string theory 
contains stable D0-branes, charged under a $U(1)$ RR symmetry.  The D0-brane  
charge can be naturally interpreted as the Kaluza-Klein momentum along an 
extra $S^1$ dimension of space, which we interpret as the extra $S^1$ of 
``noncritical M-theory'' \ncm .  It is intriguing that in the Fermi-liquid 
formulation of the eigenvalues of the matrix model, the extra $S^1$ simply 
plays the role of the angular variable on a flat two-dimensional 
plane populated by the fermions.  Noncritical M-theory is thus defined in 
terms of a large $N$, double-scaling limit of a system of $N$ nonrelativistic 
fermions in the inverted harmonic oscillator potential on this ``eigenvalue 
plane.''  We will describe the fermions in the second-quantized framework, as 
quanta of a spinless fermionic field $\Psi(t,\lambda_1,\lambda_2)$, whose 
dynamics is governed in the large $N$ limit by the nonrelativistic action 
\eqn\eedefone{S=N\int dt\,d^2\lambda\left(i\Psi^\dagger\,\frac{\p\Psi}{\p t}
-\frac{1}{2}\frac{\p\Psi^\dagger}{\p\lambda_1}\frac{\p\Psi}{\p\lambda_1}
-\frac{1}{2}\frac{\p\Psi^\dagger}{\p\lambda_2}\frac{\p\Psi}{\p\lambda_2}
+\frac{1}{2}\omega_0^2(\lambda_1^2+\lambda_2^2)\,\Psi^\dagger\Psi\right).}
\smallskip\noindent
Here $(\lambda_1,\lambda_2)$ are Cartesian coordinates on the flat 
eigenvalue plane $\R^2$, and $\omega_0$ is a fundamental frequency of the 
theory.  We will set $\omega_0=1$ throughout the paper.  

In noncritical M-theory so defined, one can find two-dimensional Type 0A and 
0B string vacua as solutions (as we will review briefly in Section~2.1).  In 
addition, the space of all solutions also includes a ``true ground state'' of 
the system, in which the lowest $N$ single-particle states have been filled 
by the $N$ available fermions, up to some distance $\mu$ from the top of 
the potential.  In the double-scaling limit, which is the semiclassical limit 
of the Fermi liquid described by \eedefone , $N$ is sent to infinity while 
the Fermi energy $\mu$ is held fixed.  $g_M=1/\mu$ plays the role of a 
coupling constant in this vacuum.  Some properties of this vacuum of 
noncritical M-theory were studied in \ncm .  

In this paper, we intend to probe this solution further, with the hope of 
learning more about the nature of its collective excitations.  For any given 
physical system, useful information about its underlying degrees of freedom 
can be revealed by exposing the system to extreme physical conditions, such 
as high temperatures, high-energy scattering, or strong fields.  This proven 
strategy was used, for example, in the early days of superstring theory:  
A series of gedanken experiments was designed (see, \eg , \refs{\grossm-
\atat}) to reveal the underlying degrees of freedom of the theory, or its 
hypothetical ``unbroken phase'' in which large underlying symmetries could 
become manifest.  In this paper, we shall apply this strategy to the vacuum 
state of noncritical M-theory, and examine it at finite temperature.%
\foot{Aspects of the high-energy behavior of noncritical M-theory will be 
studied elsewhere \appr.}

In quantum field theory, it is well-known that the high-temperature behavior 
is formally governed by a field theory in one fewer spacetime dimension, with 
the effective loop-counting coupling that scales as $T$.  Frequently, this 
naive picture is strongly modified by renormalization effects.  In string 
theory at finite temperature \refs{\alvaos-\physica,\aw,\atat}, on the other 
hand, the high-temperature behavior of one string vacuum is formally mapped by 
T-duality to the low-temperature behavior of another string vacuum in the same 
spacetime dimension.  The effective loop-counting coupling of this dual string 
theory scales as $T^2$ \aw .  In most string vacua, this naive picture is 
modified by the existence of the Hagedorn phase transition, where the 
canonical ensemble description breaks down, and one needs to resort to the 
microcanonical ensemble.  

One naturally wonders what is the high-temperature behavior of M-theory; 
in particular, is it more akin to field theory, or string theory, or neither?  
In eleven dimensions, this question again seems very difficult to answer, 
since we lack the required control over the nonsupersymmetric 
compactifications of M-theory that would formally define its thermodynamic 
ensemble in the regime of interest.%
\foot{Some papers discussing critical M-theory at finite temperature 
include \refs{\russo,\colum}; see also \oren.}
On the other hand, the same question can be addressed in noncritical M-theory 
in $2+1$ dimensions.  In Section~2, the exact partition function of the vacuum 
solution of noncritical M-theory at finite temperature will be evaluated, and 
its high temperature behavior studied.  We shall find the following:

\smallskip
\item{$\bullet$} 
The behavior of the M-theory vacuum state at high temperatures is governed by 
another, dual solution of noncritical M-theory, related to the original 
vacuum by a symmetry similar to T-duality, which noncritical M-theory inherits 
from the underlying T-duality between Type 0A and 0B strings on a compact 
Euclidean time circle.   

\smallskip
\item{$\bullet$} 
In the high-energy limit, the free energy of the system scales as $T^3$, \ie , 
in a way characteristic of a massless relativistic field theory in $2+1$ 
dimensions.  

\smallskip
\item{$\bullet$} 
In this dual M-theory solution governing the high-temperature regime, the 
effective loop-counting coupling  scales at high temperature as $T^3$.  This 
is a novel scaling behavior, unlike in field theory (${}\sim T$) or string 
theory (${}\sim T^2$).  
\smallskip
  
Our results in Section~2 will be obtained using the grand-canonical partition 
function of noncritical M-theory at finite temperature $T$.  The 
grand-canonical ensemble turns out to be equivalent, albeit in a somewhat 
subtle way involving intricacies of the double-scaling limit, to the 
canonical ensemble.  Once this equivalence is established, one can ignore 
the thermodynamic interpretation of the partition function, and simply 
interpret it as the partition function of noncritical M-theory in 
Euclidean signature, with the Euclidean time dimension compactified on a 
circle of radius $R=1/(2\pi T)$.  

This Euclidean compactification of noncritical M-theory now has {\it two\/} 
$U(1)$ isometries:  the rotation on the spatial eigenvalue plane, and the 
Euclidean time translation.  The angular dimension on the eigenvalue plane 
has been interpreted as the ``M-theory dimension'' for two-dimensional 
Type 0A strings, with the angular momentum on the plane playing the role of 
the D0-brane charge in Type 0A string theory.  In the next step, it is natural 
to ask whether the Euclidean time dimension can also be interpreted as the 
``M-theory dimension'' of some string theory.  This question will be analyzed 
in Section~3.  Somewhat surprisingly, we find evidence that this question has 
a simple answer: {\it The string theory associated with the reduction of 
noncritical M-theory along the Euclidean time circle is the closed string 
theory of the topological A-model on the resolved conifold!}  

The thermal ensemble of noncritical M-theory naturally corresponds to the 
path integral with the antiperiodic condition on the Fermi field $\Psi$ around 
the Euclidean time $t_E$.  In the study of the relation to the A-model, it is 
more natural to relax this condition, and study noncritical M-theory with 
$\Psi$ periodic up to an arbitrary phase $b$, 
\eqn\eeperiod{\Psi(t_E+2\pi R,\lambda_1,\lambda_2)=e^{ib}\Psi(t_E,\lambda_1,
\lambda_2).}
We find that this entire class of Euclidean compactifications of noncritical 
M-theory is related to the topological A-model.  In this correspondence, the 
radius $R$ of the Euclidean time circle, the coupling 
constant $g_M$ and the phase $b$ of noncritical M-theory are related to the 
A-model string coupling $g_A$ and the K\"ahler modulus $t_A$ of the resolved 
conifold by a simple duality relation, 
\eqn\eeduama{g_A=2\pi iR,\qquad t_A=\frac{2\pi R}{g_M}+ib.}
Note that what is a coupling constant on one side of the duality becomes 
a geometric parameter on the other side of the duality, and vice versa.  

This duality between noncritical M-theory and the topological A-model can 
be expected to have a wide range of interesting implications.  Some of them 
are:

\smallskip
\item{$\bullet$}  The relation between the string coupling $g_A$ and the 
radius $R$ of the M-theory circle suggests that noncritical M-theory 
can play the role of the topological M-theory \refs{\topomth,\topozth} for 
the topological strings of the A-model.  

\smallskip
\item{$\bullet$}  Noncritical M-theory provides a nonperturbative completion 
of the topological closed string theory of the A-model.  

\smallskip
\item{$\bullet$}  T-duality of noncritical M-theory (inherited from the 
underlying T-duality of two-dimensional Type 0A and 0B strings  \malsei ) 
implies an S-duality for the topological A-model.

\smallskip
\item{$\bullet$}  In combination with the Gopakumar-Vafa duality 
\refs{\gova,\govam}, our duality relates noncritical M-theory to Chern-Simons 
gauge theory on $S^3$.  

\smallskip
\item{$\bullet$}  In the melting crystal interpretation of the quantum foam 
phase of Calabi-Yau \foam , the temperature of the crystal turns out to be 
equal to the temperature of the noncritical M-theory vacuum.  The constituents 
of the crystal and the quantum foam of the Calabi-Yau appear intimately 
related to the constituent fermions of noncritical M-theory.

\newsec{Noncritical M-Theory at Finite Temperature}

Noncritical M-theory for two-dimensional Type 0A and 0B strings is defined by 
the nonrelativistic Lagrangian \eedefone\ only formally.  The proper 
definition, discussed in detail in \ncm , begins with a finite number $N$ of 
fermions, and with the inverted harmonic oscillator potential replaced with a 
more general regulating potential $V(\lambda_1,\lambda_2)$.  This potential 
includes stabilizing anharmonic terms which ensure that the spectrum is 
bounded from below.  The simplest way of mimicking such terms is to place an 
infinite wall at some distance from the origin, which cuts off the 
single-particle energy spectrum from below at some $-\Lambda$; this will be 
the regulator assumed throughout this paper.   In the large $N$ 
limit, the anharmonic terms are scaled away (or $\Lambda$ is taken to 
infinity), and the only relevant piece of the potential in this double-scaling 
limit is its inverted harmonic oscillator part.  

\subsec{Noncritical M-theory for Type 0A and 0B Strings in Two Dimensions}

Both Type 0A and 0B string vacua in two dimensions are solutions of this 
theory.  In order to see that, note that the Hamiltonian associated with 
\eedefone\ can be viewed from two complementary perspectives.  

In the polar coordinates $(\lambda,\theta)$ on the eigenvalue plane, the Fermi 
field $\Psi$ (and its conjugate momentum $\Psi^\dagger$) can 
be decomposed into eigenfunctions of the single-particle angular momentum 
operator $J$, $\Psi(\lambda,\theta)=\sum_{q\in\Z}e^{iq\theta}\Psi_q
(\lambda)$.  Consequently, the second-quantized Hamiltonian decomposes into an 
infinite sum,
\eqn\eehamsumq{\CH(\Psi,\Psi^\dagger)=\sum_{q\in\Z}\CH_q(\Psi_q,
\Psi^\dagger_q).}
Each individual term $\CH_q$ is equivalent to the Hamiltonian of the Type 0A 
string theory in the linear dilaton vacuum with RR flux $q$.  

In the Cartesian coordinates $(\lambda_1,\lambda_2)$, the single-particle 
Hamiltonian $h=h_1+h_2$ describes two decoupled oscillators.  The energy $h_2$ 
of (say) the second oscillator is a conserved quantity.  We can decompose 
$\Psi$ into a complete basis of eigenfunctions $\psi_\nu(\lambda_2)$ of 
$h_2$.  Since the eigenvalues $\nu$ of $h_2$ are continuous, we obtain an 
integral%
\foot{More precisely, for each eigenvalue of $h_2$ there are two eigenstates, 
one for each parity $\pm$ under $\lambda_2\rightarrow-\lambda_2$.  We 
implicitly include the sum over parity in our definition of the integral 
over $\nu$.}
\eqn\eedechh{\Psi(\lambda_1,\lambda_2)=\int d\nu\,\Psi_\nu(\lambda_1)\,
\psi_\nu(\lambda_2).}
As a result, the second-quantized Hamiltonian $\CH$ decomposes into an integral
over $\nu$ of a one-parameter family of decoupled Hamiltonians $\CH_\nu$, each 
of which depends only on a single second-quantized field 
$\Psi_\nu(\lambda_1)$.  Each member of this family of Hamiltonians is 
essentially equivalent to the Hamiltonian of two-dimensional Type 0B string 
theory in Fermi liquid representation.  

Having understood these facts about noncritical M-theory, it is now clear how 
the Type 0A and 0B vacua can be constructed as exact solutions of this 
theory.  The Type 0A linear dilaton vacuum with RR flux $q$ is given simply 
by the state of the $2+1$ dimensional noncritical M-theory in which all 
single-particle states with angular momentum $J=q$ are occupied up to some 
(double scaled) fermi energy $\mu=-N\epsilon_F$, while all fermion states with 
$J\neq q$ are empty.  The physics of this solution is {\it exactly 
equivalent\/} to the physics of the linear dilaton ground state of the Type 
0A theory in the Fermi liquid representation.  In particular, since the Fermi 
sea of this state is empty in all sectors with $J\neq q$, all excitations in 
those sectors are infinitely energetic from the perspective of the Type 0A 
ground state, and they decouple in the double-scaling limit.  The physics of 
the remaining excitations is then isomorphic to that of Type 0A string 
theory in the Fermi liquid picture.  

Similarly, the Type 0B linear dilaton vacuum corresponds to the Fermi sea in 
which the $N$ fermions occupy (up to the Fermi energy $\mu$) only the 
single-particle states whose eigenvalue of $h_2$ is equal to some fixed 
$\nu$, while all single-particle states whose eigenvalue $h_2\neq\nu$ are 
empty.  In the double-scaling limit, all excitations with $h_2\neq\nu$ are 
infinitely energetic and decouple.  The physics of the remaining finite-energy 
excitations of this solution is precisely equivalent to the Fermi liquid 
representation of the Type 0B string theory.  In the Type 0A and 0B string 
vacua of noncritical M-theory, the fundamental frequency $\omega_0$ plays the 
role of $1/\sqrt{2\alpha'}$ \ncm .  

In addition to reproducing such known two-dimensional string vacua, 
noncritical M-theory has other interesting solutions.  The one of primary 
interest to us is the ``true'' ground state of the theory, in which the $N$ 
available fermions simply occupy {\it the lowest $N$ single-particle states\/} 
of the theory, irrespective of what other quantum numbers they may 
carry (and with the Fermi energy $\mu$ held fixed in the large $N$ limit).  
\fig{The ground state of noncritical M-theory in the Fermi liquid 
representation.  The inverted harmonic oscillator potential is rotationally 
invariant on the eigenvalue plane, with the second dimension suppressed in 
the figure.  All single-particle states are occupied from the cutoff up to the 
top of the Fermi sea, some distance $\mu$ below the top of the 
potential.}{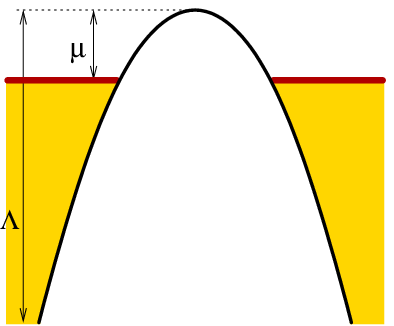}{2truein}
In this simplified context of noncritical M-theory, it is this natural vacuum 
state which is perhaps the closest analog of the eleven-dimensional vacuum of 
full M-theory.  Various properties of this vacuum state of noncritical 
M-theory were studied in \ncm , and we will review some of them in 
Section~2.5 below. 

It is also worthwhile to point out that families of static solutions 
interpolating between the M-theory vacuum and the Type 0A or 0B string 
theory can be constructed.  Starting from Type 0A (or 0B) solutions as 
described above, one can raise the Fermi surface in sectors with $J\neq q$ 
(or $h_2\neq\nu$) to some finite distance below the top of the potential, 
making the excitations in those sectors finitely energetic.  This change in 
the Fermi surface represents a hidden deformation parameter of two-dimensional 
Type 0A and 0B string theory.

\subsec{The Grand Canonical Ensemble}

We find it convenient to analyze the thermodynamics of noncritical M-theory 
using the grand canonical ensemble of the fermions.  In this ensemble, the 
central object of our interest will be the thermodynamic potential 
$\Gamma(\beta,\mu)$, which is a function of the inverse temperature 
$\beta=1/T$ and the (double-scaled) chemical potential $\mu$ associated with 
the conserved number of fermions $N$. $\Gamma(\beta,\mu)$ is defined in terms 
of the grand canonical partition function $\CZ_M(\beta,\mu)$ of the 
noncritical M-theory vacuum, 
\eqn\eefreformal{\Gamma(\beta,\mu)=\frac{1}{\beta}\log\CZ_M=\frac{1}{\beta}
\int_{-\infty}^{\infty}d\xi\,\rho(\xi)\log\left(1+e^{\beta(\mu-\xi)}
\right).}
Here $\log\CZ_M$ has been related to the single-particle density of states 
$\rho(\xi)$ at zero temperature and at energy $\xi$, a quantity studied in 
detail in \ncm .  

In the thermodynamic limit, which is a part of the double-scaling limit, this 
ensemble is equivalent to the canonical ensemble provided the fluctuations in 
the number of particles are suppressed, a condition that we will return to in 
Section~2.4 below.  Potentially, the canonical ensemble in string theory is 
known to suffer from several instabilities, including the Jeans instability 
common to all gravitating systems, and the stringy instability due to the 
Hagedorn phase transition.  These destabilize the canonical ensemble and 
require the full power of the microcanonical ensemble.  Such instabilities do 
not occur in Type 0A and 0B string theories in two dimensions, and we will see 
that they do not occur in noncritical M-theory either.  The microcanonical, 
canonical and grand canonical ensembles are all equivalent, but the argument 
is surprisingly subtle and involves the double-scaling limit in a nontrivial 
way.  

In \ncm , an exact formula for the density of states $\rho(\xi)$ in the 
M-theory vacuum was derived, 
\eqn\eedensex{\eqalign{\rho(\xi)&=\frac{1}{4\pi}\Re\int_0^\infty d\tau
e^{-i\xi\tau}\,\frac{1}{\sinh^2(\tau/2)}\cr
&\qquad{}=-\frac{\xi}{2\tanh(\pi\xi)}+\frac{\Lambda}{2}.\cr}}
The integral representation for the density of states in \eedensex\ is 
divergent, but the divergence is rather mild,  with the divergent piece 
independent of $\mu$ and resulting in a simple dependence of $\rho(\xi)$ 
on the cutoff $\Lambda$.  In the double scaling limit, $\Lambda$ is scaled 
to infinity with $N$.  As in \ncm , we only keep the leading dependence 
on $\Lambda$, dropping all terms $\CO(1/\Lambda)$.  
\fig{The density of states \eedensex\ as a function of the single-particle 
energy $\xi$ of the fermions.  The density is invariant under $\xi\rightarrow 
-\xi$.  Notice also the importance of the nonuniversal $\xi$-independent 
constant in \eedensex : if we ignored this $\Lambda$-dependent term, the 
density of states would be negative.}{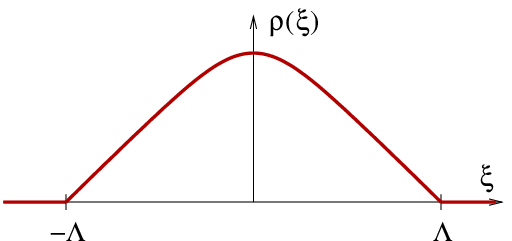}{3truein}

Similarly, the formal expression \eefreformal\ for $\Gamma$ is divergent.  
In most of our arguments, we will concentate only on the universal part of 
all the thermodynamic variables.  Thus, in order to eliminate all dependence 
on $\Lambda$ and isolate the universal information in $\Gamma$, we use a 
trick standard in noncritical string theory, and evaluate
\eqn\eefreexder{\frac{\p^3\Gamma}{\p\mu^3}=-\beta^2\int_{-\infty}^\infty d\xi\,
\frac{\p\rho(\xi)}{\p\xi}\,\frac{1}{4\cosh^2\left[\beta (\xi-\mu)/2\right]},}
which is cutoff-independent.  We will return to the dependence of the 
thermodynamic quantities on the cutoff in the next subsection.  

Using the integral representation \eedensex\ for $\rho(\xi)$ and performing 
the integral over $\xi$ by residues, we get
\eqn\eeintgam{\eqalign{\frac{\p^3\Gamma}{\p\mu^3}&=-\frac{1}{4\pi}
\int_{-\infty}^\infty d\xi\,\Re\int_0^\infty d\tau\,e^{i\xi\tau}
\frac{\tau^2}{\sinh^2(\tau/2)}\,\frac{1}{1+e^{2\pi R(\xi-\mu)}}\cr
&=-\frac{1}{2\pi}\,\Im\int_0^\infty d\tau\,e^{i\mu\tau}
\frac{\tau/2}{\sinh^2(\tau/2)}\,\frac{\tau/(2R)}{\sinh{[\tau/(2R)]}}.\cr}}
This integral representation for the thermodynamic potential of the M-theory 
vacuum is similar to the corresponding expressions in noncritical string 
theory  \grosskl\ (see also \refs{\nakarev,\alexrev} for recent reviews).  
Note that -- unlike for example in noncritical bosonic string theory -- the 
M-theory formula does not exhibit any manifest form of self T-duality.  
However, as we will see later, M-theory does inherit a form of T-duality from 
the underlying T-duality relating Type 0A and 0B strings with RR flux.  

\subsec{Terms of Low Order in $\mu$}

The integral formula \eeintgam\ is convergent, but it determines 
$\Gamma(\beta,\mu)$ only up to a second order polynomial in $\mu$, 
\eqn\eepolynund{\Gamma_0(\beta)+\Gamma_1(\beta)\,\mu+\Gamma_2(\beta)\,
\frac{\mu^2}{2}.}
The coefficients $\Gamma_i(\beta)$ are cutoff-dependent, and divergent in the 
double-scaling limit.  However, before we dismiss them as non-universal, 
we should note that they do contain some valuable physical information.  
Imagine, for example, that one is interested in the behavior of our system 
at finite temperature, and for zero chemical potential $\mu=0$. 
In such circumstances, the thermodynamic potential $\Gamma$ would be given 
entirely by the lowest term $\Gamma_0(\beta)$ in \eepolynund\ (at least if we 
assume analyticity at $\mu=0$).  

With such applications in mind, it is useful to separate the universal 
information in $\Gamma_i(\beta)$ from the nonuniversal terms.  Even though 
$\Gamma_i(\beta)$ are all divergent as we take the cutoff to infinity, we can 
take additional derivatives with respect to $\beta$, and obtain
\eqn\eegamzero{\frac{\p^2(\beta\Gamma_0)}{\p\beta^2}=\frac{1}{4}
\int_{-\infty}^\infty d\xi\,\rho(\xi)\frac{\xi^2}{\cosh^2(\beta\xi/2)}
=\frac{\Lambda}{2\beta}-\frac{1}{8}\int_{-\infty}^\infty d\xi\,
\frac{\xi^3}{\tanh(\pi\xi)\,\cosh^2(\beta\xi/2)}.}
In the last step, we have used the expression \eedensex\ for $\rho(\xi)$, and 
evaluated the convergent integral that multiplies the $\Lambda$ term.  The 
remaining integral is now also convergent, and \eegamzero\ determines the 
dependence of $\beta\Gamma_0(\beta)$ on $\beta$ up to a (possibly divergent) 
nonuniversal polynomial linear in $\beta$.  

Similarly, 
\eqn\eegamone{\frac{\p\Gamma_1}{\p\beta}=-\frac{1}{4}
\int_{-\infty}^\infty d\xi\,\rho(\xi)\frac{\xi}{\cosh^2(\beta\xi/2)}=0.}
This integral vanishes because $\rho(\xi)$ is an even function of $\xi$, 
implying that $\Gamma_1$ is independent of $\beta$, and equal to a 
(possibly divergent) nonuniversal constant.  

As to the $\Gamma_2$ term, its divergence comes completely from the linear 
divergence of $\rho$, and we can evaluate it without taking any additional 
derivatives,
\eqn\eegamtwo{\Gamma_2=\beta\int_{-\infty}^\infty d\xi\,\rho(\xi)
\frac{1}{\cosh^2(\beta\xi/2)}=\frac{2\pi^2\Lambda}{3\beta^2}-
\frac{\beta}{2}\int_{-\infty}^\infty d\xi\,\frac{\xi}{\tanh(\pi\xi)\,
\cosh^2(\beta\xi/2)}.}
Despite being non-universal, the leading $\Lambda$-dependent terms in 
\eegamzero\ and \eegamtwo\ will play an important role in our discussion of 
the equivalence of various thermodynamic ensembles in the next subsection.  

\subsec{Remarks on Thermodynamic Stability}

In order to substantiate the discussion of the grand canonical ensemble, 
we must examine whether this ensemble is well-defined, and if so, whether 
it is equivalent to the canonical and the microcanonical ensemble.  In order 
to test that, we will test the validity of the fluctuation-dissipation 
theorem.  This theorem relates the fluctuations in the mean energy $\langle E
\rangle$ and the mean number of particles $\langle N\rangle$ to thermodynamic 
quatities: the specific heat and the isothermal compressibility.  In terms 
of $\Gamma$, we have
\eqn\eethermquant{\langle N^2\rangle-\langle N\rangle^2=\frac{1}{\beta}\,
\frac{\p^2\Gamma}{\p\mu^2}.}
Similarly, the specific heat $C_V$ is the measure of the energy fluctuations, 
\eqn\eeflucten{\langle E^2\rangle-\langle E\rangle^2=\frac{C_V}{\beta^2}\equiv
\frac{\p^2(\beta\Gamma)}{\p\beta^2}.}
First of all, in order for the grand canonical ensemble to be well-defined, 
both of these quantities need to be positive.  Using \eethermquant\ and 
\eeflucten\ together with our evaluation of the leading divergent terms in 
\eegamzero\ and \eegamtwo , we deduce that
\eqn\eethermm{\langle N^2\rangle-\langle N\rangle^2=\frac{\Lambda}{2\beta}
-\frac{1}{8}\int_{-\infty}^\infty d\xi\,\frac{\xi}{\tanh(\pi\xi)\,
\cosh^2[\beta(\xi-\mu)/2]},}
and 
\eqn\eefluctenm{\langle E^2\rangle-\langle E\rangle^2=\frac{\Lambda}{2\beta}
\left(\frac{\pi^2}{3\beta^2}+\mu^2\right)
-\frac{1}{8}\int_{-\infty}^\infty d\xi\,\frac{\xi^3}{\tanh(\pi\xi)\,
\cosh^2[\beta(\xi-\mu)/2]}.}
As we see, if we discarded the nonuniversal $\Lambda$-dependent pieces in 
\eethermm\ and \eefluctenm\ and kept only the universal terms dependent on 
$\mu$, both quantities would be negative, suggesting a possible thermodynamic 
instability of the canonical as well as the grand canonical ensemble.  The 
proper behavior of the ensembles is restored by the leading nonuniversal term 
proportional to the cutoff $\Lambda$ in \eethermm\ and \eefluctenm .  These 
terms are always positive, and stabilize both ensembles in the double-scaling 
limit.   

Having established the stability of the ensembles, it is also straightforward 
to verify that the fluctuations of the mean values $\langle N\rangle$ and 
$\langle E\rangle$ are small in the thermodynamic limit, \ie , that 
\eqn\eesupfl{\frac{\langle N^2\rangle-\langle N\rangle^2}{\langle N\rangle^2}
\quad{\rm and}\quad \frac{\langle E^2\rangle-\langle E\rangle^2}{\langle 
E\rangle^2}}
vanish.  In the thermodynamic limit, we take $\Lambda\rightarrow\infty$ while 
keeping $\mu$ and $\beta$ fixed, the denominators in \eesupfl\ indeed diverge 
faster than \eethermm\ and \eefluctenm , and all three ensembles are 
equivalent.  (In view of this, we can now drop the brackets in 
$\langle N\rangle$ and $\langle E\rangle$, and denote them simply by $N$ and 
$E$.)

\subsec{Expansion in the Powers of the Coupling $1/\mu$}

In the matrix models of two-dimensional string theory, the inverse Fermi 
energy $1/\mu$ plays the role of the string coupling, $g_s=1/\mu$.  The exact 
partition function can be expanded as an asymptotic series%
\foot{Throughout this paper, we use the symbol ``$\approx$'' to denote exact 
asymptotic expansions of various formulas.}
in the powers of $g_s$,

\eqn\eestrpartf{F_{\rm string}\approx \sum_{h=0}^\infty F_h(R)\,g_s^{2h-2}}
and compared to the perturbative string-theory calculation.  

In the vacuum of noncritical M-theory, similarly, the inverse Fermi energy 
plays the role of an ``M-theory coupling constant,'' which we denote by%
\foot{Folklore says that M-theory in eleven dimensions contains no 
dimensionless couplings.  However, on a nontrivial background geometry with a 
characteristic length scale $\CR$, a natural dimensionless coupling emerges as 
$\CR$ measured in Planck units.  We believe that $g_M$ should be thought of as 
a dimensionless coupling of such a geometric origin.  In fact, a geometric 
interpretation for $g_M$ will indeed be found in Section~3.}
\eqn\eqngm{g_M=1/\mu.}
One can expand physical quantities in the powers of $g_M$, and contrast the 
behavior of this expansion with the expansion \eestrpartf\  in two-dimensional 
string theory.  In \eestrpartf , the higher genus coefficients $F_h$, with 
$h>1$, are universal functions of the geometric moduli, such as the radius $R$ 
of the Euclidean time circle.  In addition, the lowest genus terms $F_0$ and 
$F_1$ also depend logarithmically on $\mu$ and on the cutoff $\Lambda$.  
Moreover, if open strings are present, odd powers of $g_s$ can also appear 
in \eestrpartf .  In this sense, the string perturbation expansion is 
naturally an expansion in ``halves of loops.''  More precisely, we wish to 
interpret \eestrpartf\ as a semiclassical expansion of an effective action 
$S_{\rm eff}/\hbar$ in powers of $\hbar$.  The power of 
$\hbar$ counts by definition the number of loops.  The leading-order term 
in \eestrpartf\ is by definition classical, and therefore of order $1/\hbar$.  
The one-loop term in $S_{\rm eff}/\hbar$ is independent of $\hbar$.  Thus, we 
see the well-known fact that the role of $\hbar$ (or the ``loop-counting 
parameter'') in string theory is effectively played by $g_s^2$.  Since the 
disk diagram would be proportional to $1/g_s$, this diagram effectively 
contributes at ``one-half-loop'' order compared to the classical term 
originating from the sphere $1/g_s^2$.  

In the case of the vacuum energy of the M-theory vacuum state at zero 
temperature, this expansion was studied in detail in \ncm , and found to 
take the form
\eqn\eempartf{\CF_M\approx \sum_{n=0}^\infty \CF_n\,g_M^{n-3}.}
This expansion exhibits several intriguing features \ncm : 

\smallskip
\item{$\bullet$} 
The leading term in \eempartf\ is of order $1/g_M^3$.  In the semiclassical 
expansion, this will be the leading, classical term, implying that the role 
of the Planck constant in noncritical M-theory is played by 
$\hbar\sim g_M^3$.  Since the subleading corrections found in \ncm\ are 
integer powers of $g_M$, the natural expansion is in ``thirds of loops,'' or 
$\hbar^{1/3}\sim g_M$.  This is intriguingly reminiscent of heterotic 
M-theory \hetm\ in eleven dimensions. 

\smallskip
\item{$\bullet$} Unlike in two-dimensional Type 0A/0B string theory, where 
all orders in \eestrpartf\ are nontrivial, it turns out that the perturbative 
expansion \eempartf\ in noncritical M-theory terminates at one-loop.   All 
coefficients $\CF_n$ with $n>3$, \ie, all terms in \eempartf\ proportional 
to positive powers of $g_M$, are identically zero!  

\smallskip
\item{$\bullet$} The instanton corrections to \eempartf\ were also determined 
in \ncm .  They have characteristic nonperturbative weights
\eqn\eeinstw{\exp\left(-\frac{2\pi n}{g_M}\right)\sim
\exp\left(-\frac{2\pi n}{\hbar^{1/3}}\right),}
with $n$ any positive integer.  According to a classic argument \shenker , 
nonperturbative effects in string theory are stronger than in field theory,  
since they scale as $\exp(-C/\hbar^{1/2})$ compared to the field-theory 
behavior $\exp(-C/\hbar)$.  By the same argument, \eeinstw\ indicates that 
nonperturbative effects in noncritical M-theory are even stronger than in 
string theory.  

\smallskip
\item{$\bullet$} For each $n$, the instanton term \eeinstw\ is multiplied by 
a perturbative expression similar to \eempartf .  This expression starts at 
one-third loop order $\hbar^{1/3}$ (\ie , the leading, classical term is 
absent), and is also one-loop exact.
\smallskip

It is easy to show that the one-loop exactness of the asymptotic expansion 
\eempartf\ persists at finite temperature as well.  The expression \eeintgam\ 
for the (third derivative of) the thermodynamic potential can be expanded 
in the powers of $1/\mu$ at fixed $R$ (after changing the integration variable 
to $\sigma=\mu\tau$ as in \ncm ).  This  yields
\eqn\eelargemuexp{\frac{\p^3\Gamma}{\p\mu^3}\approx-\frac{1}{\pi}\int_0^\infty
\frac{d\sigma}{\sigma}\sin(\sigma)\left(1+
\sum_{n=1}^\infty f_n(R)\frac{\sigma^{2n}}{\mu^{2n}}\right)e^{-\epsilon
\sigma}.}
We do not even need to determine the precise form of the $R$-dependent 
coefficients $f_n(R)$, because the integrals that they multiply all vanish 
identically, with the exception of the leading constant term.  
Thus, the $1/\mu$ expansion \eelargemuexp\ terminates at the lowest order,
\eqn\eelargemuexpex{\frac{\p^3\Gamma}{\p\mu^3}\approx-\frac{1}{2}
+{\rm nonperturbative\ terms},}
which in turn implies that
\eqn\eegammalmuex{\Gamma(\mu,R)\approx -\frac{1}{12}\mu^3+\CF_1(R)\,\mu^2+
\CF_2(R)\,\mu+\CF_3(R)+{\rm nonperturbative\ terms}.}
In addition to their dependence on the temperature, the subleading terms 
$\CF_i$, $i=1,2,3$ may also depend on the nonuniversal cutoff $\Lambda$.  

We have not checked explicitly the behavior of the perturbative expansions 
near the instanton corrections, but we expect them to continue to be 
one-loop exact at finite temperature as well.   For the special case of $R=1$, 
this will be explicitly verified in Section~2.9.

\subsec{The Low-Temperature Expansion}

We shall now analyze the thermodynamic behavior of noncritical M-theory in 
the regimes of low and high temperature.  

First, we expand $\Gamma$ as an asymptotic expansion at low temperature 
$T=1/(2\pi R)$ at fixed $\mu$:
\eqn\eelowtemp{\eqalign{\frac{\p^3\Gamma}{\p\mu^3}&\approx-\frac{1}{4\pi}
\int_0^\infty d\tau\,\sin(\mu\tau)\,\frac{\tau}{\sinh^2(\tau/2)}
\left(1-\sum_{k=1}^\infty\frac{2(2^{2k-1}-1)B_{2k}}{(2k)!}\left(\frac{\tau}{2R}
\right)^{2k}\right)\cr
&=-\frac{1}{4\pi}\sum_{k=0}^\infty\frac{1}{R^{2k}}\,
\frac{(1-2^{1-2k})(-1)^{k-1}B_{2k}}{(2k)!}\frac{\p^{2k}}{\p\mu^{2k}}
\int_0^\infty d\tau\,\frac{\tau}{\sinh^2(\tau/2)}\,\sin(\mu\tau).\cr}}
$B_m$ are the Bernoulli numbers (see \ncm\ for our conventions).  The 
remaining integral in \eelowtemp\ can be evaluated and the series formally 
resummed into the following compact expression, 
\eqn\eecompex{\frac{\p^3\Gamma}{\p\mu^3}\approx-\frac{R\left(\frac{1}{2R}
\,\frac{\p}{\p\mu}\right)^2}{\sin\left(\frac{1}{2R}\,\frac{\p}{\p\mu}\right)}
\,\frac{\mu}{\tanh(\pi\mu)},}
which should be interpreted in the sense of an asymptotic expansion in 
$1/R$. Integrating \eelowtemp\ once, we obtain
\eqn\eelowtempsec{\frac{\p^2\Gamma}{\p\mu^2}\approx
-\frac{1}{2}\sum_{k=0}^\infty\frac{1}{R^{2k}}\,\frac{(1-2^{1-2k})
(-1)^{k-1}B_{2k}}{(2k)!}\frac{\p^{2k}}{\p\mu^{2k}}\,
\frac{\mu}{\tanh(\pi\mu)}+\frac{\Lambda}{2}.}

This result is in accord with the following general observation.  The number 
of particles $N$ is related to $\Gamma$ via
$$N=\frac{\p\Gamma}{\p\mu}.$$
Taking a derivative of both sides, we obtain
$$\rho_{\rm eff}(\mu,\beta)\equiv\frac{\p N}{\p\mu}=\frac{\p^2\Gamma}{\p\mu^2},
$$
where $\rho_{\rm eff}(\mu,\beta)$ is the effective density of states at finite 
temperature.  It is reassuring to see that the leading term in \eelowtempsec\ 
is indeed the zero-temperature density of states $\rho(\mu)$, and reproduces 
the zero-temperature result of \ncm .   

Perhaps the most interesting feature of the low-temperature expansion 
\eelowtempsec\ is the fact that all dependence on $T$ in \eelowtempsec\ falls 
off rapidly as $\mu>1$.  This suggests a radical reduction of the effective 
number of degrees of freedom in the theory at small values of the M-theory 
coupling $g_M$.  This is compatible with the one-loop exactness of the 
weak-coupling expansion observed in \ncm\ and in the previous subsection.  
Both of these features suggest that the theory becomes effectively topological 
for small values of $g_M$, with a smooth crossover to a more dynamical regime 
at strong coupling.  

\subsec{The High-Temperature Expansion}

At high temperature, we can expand the thermodynamic potential $\Gamma$ in 
the powers of $R$, 
\eqn\eehightemp{\frac{\p^3\Gamma}{\p\mu^3}\approx\frac{R}{2\pi}
\int_0^\infty \frac{d\sigma\,\sigma}{\sinh(\sigma/2)}
\sum_{m=0}^\infty\frac{(-1)^m}{(2m+1)!}
\sum_{k=0}^\infty\frac{(2k-1)B_{2k}}{(2k)!}\mu^{2m+1}(\sigma R)^{2m+2k}.}
First, we wish to isolate the behavior of the system as $T\rightarrow
\infty$.  At high temperature, the system can be expected to become 
effectively classical, and we must find the correct way of identifying this 
classical limit.   Holding $\mu$ while taking the high-temperature limit 
$R\rightarrow 0$ in \eehightemp\ would lead to 
\eqn\eehightemp{\frac{\p^3\Gamma}{\p\mu^3}\approx-\frac{\pi\mu R}{2}+\CO(R^3).}
Hence, in this naive limit of infinite temperature with $\mu$ fixed, 
$\p^3\Gamma/\p\mu^3$ would vanish.  In order to obtain a sensible classical 
limit at high temperatures, we must instead scale $\mu$ with $R$ such that 
\eqn\eemudual{\mudual\equiv R\,\mu}
is held fixed as $R\rightarrow 0$.  This leads to a nontrivial 
high-temperature behavior, 
\eqn\eeleads{\eqalign{\frac{\p^3\Gamma}{\p\mu^3}&\approx-\frac{1}{2\pi}
\int_0^\infty d\sigma\,\sin(\mudual\sigma)
\left(\frac{1}{\sigma}+\CO(R^2)\right)\frac{\sigma}{\sinh(\sigma/2)}\cr
&\qquad\qquad{}=-\frac{1}{2}\tanh(\pi\mudual)+\CO(R^2),\cr}}
in which all orders in $\mudual$ contribute at the leading order in $T$.  
Thus, the proper classical limit of noncritical M-theory at high 
temperature involves holding $\mudual$ fixed as $T$ goes to infinity.  

We are now ready to compare the high-temperature behavior of noncritical 
M-theory to the high-temperature behavior in field and string theory.%
\foot{A nice discussion of the high-temperature classical limit in string 
theory and field theory can be found, for example, in the classic paper by 
Atick and Witten \aw .}

In quantum field theory, consider for illustration the case of Yang-Mills 
gauge theory.  The high-temperature limit is formally governed by an effective 
field theory in one fewer spacetime dimension, with effective loop-counting 
coupling given by $g_{\rm YM,\ eff}^2=g_{\rm YM}^2T$.  In order to obtain the 
proper classical limit at high temperatures, we must keep $g_{\rm YM,\ eff}^2$ 
fixed while taking $T$ to infinity.  Of course, in most field theories, this 
naive picture will be strongly modified by renormalization effects and 
infrared divergences.  

String theory at high temperature does not undergo an effective dimensional 
reduction, and is instead formally governed by a T-dual solution in the same 
dimension at the dual temperature.  The effective loop-counting coupling of 
the dual theory is given in terms of the original string coupling by 
$g_s^2T^2$.  In order to obtain a consistent classical limit at high 
temperatures, $g_s^2T^2$ must be held fixed as $T$ goes to infinity.  In this 
limit, as was pointed out in \aw , the free energy of the system scales as 
$T^2$, \ie , as in a conformal field theory in $1+1$ spacetime dimensions.  
In critical string theory, this formal picture is strongly modified by the 
Hagedorn phase transition.  

With this picture in mind, it is natural to ask the following questions in 
noncritical M-theory: 

\item{(1)} What is the effective theory that governs the high-temperature 
behavior of the noncritical M-theory vacuum?

\item{(2)} What is the leading behavior of the thermodynamic potential (or  
the free energy) at high temperature?

\item{(3)} In the effective theory governing the high-temperature behavior,  
how does the effective coupling scale with $T$?  

We start answering these questions by isolating the leading high-temperature 
behavior of the thermodynamic potential $\Gamma$.  As we have seen, $\mudual$ 
is held fixed in the high-temperature limit.  Upon rewriting \eeleads\ in 
terms of $\mudual$ and $T$, we obtain
\eqn\eedereff{\frac{\p^3\Gamma}{\p\mudual^3}=(2\pi T)^3
\left(-\frac{1}{2}\tanh(\pi\mudual)\right)+\ldots,}
where the ``$\ldots$'' denote all terms subleading in $T$.  Thus, at high 
temperatures, the most divergent term in the thermodynamic potential (and, 
consequently, also in the free energy) scales as
\eqn\eescalehight{\Gamma(T,\mudual)\sim T^3.}
This high-temperature scaling of $\Gamma$ as $T^3$ is a behavior 
characteristic of a relativistic massless field theory in $2+1$ spacetime 
dimensions.  This observed high-temperature behavior of the noncritical 
M-theory vacuum lends further support to the arguments of \ncm\ that the 
collective excitations of the Fermi surface in this vacuum are effectively 
described by a bosonic collective field in $2+1$ dimensions.

At very high temperatures, the physics of the noncritical M-theory vacuum is 
governed by a dual effective theory.  We claim that this dual theory is 
another solution of noncritical M-theory in $2+1$ dimensions, 
effectively at zero temperature, and with the effective M-theory coupling 
constant given by 
\eqn\eegdual{\widetilde g_M=1/\mudual.}
This claim will be further substantiated by a duality argument in the next 
subsection.  Here, we present first evidence for this claim.  \eedereff\ 
implies that the partition function $\widetilde\Gamma(\mudual)$ of 
the effective dual theory that governs the classical limit at high temperature 
should satisfy
\eqn\eepfeff{\frac{\p^3\widetilde\Gamma}{\p\mudual^3}\sim-\frac{1}{2}\tanh(
\pi\mudual).}
This can be asymptotically expanded in the powers of $\widetilde g_M$,
\eqn\eepfeffexp{\frac{\p^3\widetilde\Gamma}{\p\mudual^3}\sim-\frac{1}{2}
+ {\rm nonperturbative\ terms}.}
As a result, the leading term in the asymptotic expansion of 
$\widetilde\Gamma(\mudual)$ in the powers of $1/\mudual$ is proportional to 
$\mudual^3$, and the asymptotic expansion is one-loop exact.  As we saw 
in \ncm , and again in Section~2.5 of this paper, this behavior is indeed 
a hallmark of noncritical M-theory in $2+1$ dimensions.   

We can gain further insight into the nature of this dual solution of 
M-theory as follows.  As in any other solution of the double-scaled Fermi 
liquid theory, the right-hand side of \eepfeff\ should be 
equal to the derivative of the effective density of states $\widetilde\rho 
(\mudual)$ of this solution.  Thus, integrating \eeleads\ once, we obtain -- 
up to a nonuniversal integration constant -- the effective density of states 
of the dual solution that represents the high-temperature limit of the 
M-theory vacuum,
\eqn\eedensihigh{\widetilde\rho(\mudual)\sim-\frac{1}{2\pi}\log\cosh(\pi
\mudual).}
Thus, even though this density of states exhibits the leading asymptotic 
behavior characteristic of noncritical M-theory, it is strictly distinct from 
the density of states \eedensex\ of the original vacuum at zero temperature, 
demonstrating that noncritical M-theory is not self-dual under the exchange of 
the high and low temperature regions.  

Since the effective dual theory at high temperatures is another solution 
of noncritical M-theory, our arguments from Section~2.5 imply that the 
effective loop-counting coupling in this solution is given by $\widetilde 
g_M^3$.  This is related to the coupling $g_M^3$ of the original vacuum by 
\eqn\eescales{\widetilde g_M^3=(2\pi)^3\,g_M^3\,T^3.}
Thus, the effective loop-counting coupling in the high-temperature limit 
scales as $T^3$.  This behavior seems to be a novel signature of M-theory, 
to be contrasted with the behavior in quantum field theory as well as string 
theory, as reviewed above.  It would be very interesting to see whether a 
similar $\sim T^3$ scaling can also be found in full eleven-dimensional 
M-theory and its compactifications.  

To summarize, the answers to the three questions about the high temperature 
behavior of noncritical M-theory are as follows:

\item{(1)} In the high temperature limit, the M-theory vacuum is effectively 
described by another solution of noncritical M-theory, effectively at zero 
temperature, and with the density of states given by \eedensihigh .

\item{(2)} The free energy in the noncritical M-theory vacuum scales 
at high temperatures as $T^3$, \ie , as in a massless field theory in $2+1$ 
dimensions.  

\item{(3)} The scaling \eescales\ of the effective coupling constant with $T$ 
is unlike in field or string theory, suggesting that in this phase, 
noncritical M-theory is not (manifestly) equivalent to either.  

\subsec{Effective M-Theory at High Temperature from T-Duality}

In the previous subsection, we found signs indicating that the 
high-temperature limit of the noncritical M-theory vacuum is effectively 
described by another solution of noncritical M-theory in $2+1$ dimensions.  
However, the exact nature of that solution was left somewhat obscure.  Now 
we will use T-duality of Type 0A and 0B string theories to identify this 
dual solution of noncritical M-theory.  

As was reviewed in Section~2.1, the Hamiltonian of noncritical M-theory in 
the second-quantized fermionic representation can be expressed as an infinite 
sum of sectors with all possible integer values $q$ of the angular momentum
$J$ on the plane.  For each fixed $q$, the Hamiltonian is equivalent to that 
of the linear dilaton vacuum of Type 0A string theory with RR flux equal to 
$q$.  In the vacuum state of noncritical M-theory, all available states are 
filled by fermions up to some common value $\mu$ of the Fermi energy in all 
sectors independently of the value of $q$.  Using this decomposition, many 
physical quantities of the noncritical M-theory vacuum can be formally 
evaluated as infinite sums of contributions from Type 0A vacua with all 
possible values of the RR flux $q$.  For example, the vacuum energy of the 
vacuum solution equals the sum of vacuum energies of Type 0A vacua of all 
integer values of $q$, all filled up to the common value of the Fermi energy 
$\mu$.  Similarly, the density of states \eedensex\ in M-theory can be 
evaluated as a sum over $q$ of densities of states in Type 0A vacua at fixed 
$q$,
\eqn\eedenssum{\rho(\xi)=\frac{1}{4\pi}\Re\int d\tau
e^{-i\xi\tau}\,\frac{1}{\sinh^2(\tau/2)}
=\sum_{q\in\Z}\frac{1}{2\pi}\Re\int d\tau e^{-i\xi\tau}\,\frac{1}{\sinh(\tau)}
e^{-|q|\tau}.}

Consider now noncritical M-theory at finite temperature $T$.  Using 
\eedenssum , the thermodynamic potential $\Gamma$ can be written as an 
infinite sum of contributions from sectors of fixed angular momentum $q$.  
Recalling the integral formula \eeintgam , we can write
\eqn\eeintgsum{\eqalign{2\pi R\cdot\Gamma=
&=-\frac{1}{4}\Re\int_0^\infty \frac{d\tau}{\tau}\,e^{i\mu\tau}
\frac{1}{\sinh^2(\tau/2)}\,\frac{1}{\sinh{[\tau/(2R)]}}\cr
&\quad{}=-\frac{1}{2}\sum_{q\in\Z}\Re\int_0^\infty\frac{d\tau}{\tau}
\,e^{i\mu\tau}
\frac{1}{\sinh(\tau)}\,\frac{1}{\sinh{[\tau/(2R)]}}\,e^{-|q|\tau}.\cr}}
Each contribution of fixed $q$ in \eeintgsum\ is equivalent to $2\pi R\cdot 
\Gamma$ evaluated in Type 0A theory with RR flux $q$, at temperature $T$.  

In the canonical ensemble, Type 0A theory at temperature $T$ is described by 
the solution with the Euclidean time compactified on a circle of radius 
$R=1/(2\pi T)$.  It was shown in \malsei\ that this Euclidean solution of 
Type 0A string theory is T-dual to a Type 0B solution at the dual values of 
the radius and of the inverse string coupling,%
\foot{For some earlier work on T-duality in two-dimensional Type 0A and 0B 
string vacua, see \refs{\grossw-\ulf}.}
\eqn\eeoaobdual{\widetilde R=\frac{1}{2R}, \qquad \mudual=R\,\mu.}
In addition, this dual Type 0B solution exhibits a nonzero value $\nu$ of the 
one-form RR flux, given by \malsei\ 
\eqn\eedrrflux{\nu=\frac{iq}{2\widetilde R}.}
At finite $\widetilde R$, the unit of the Type 0B RR flux is quantized, 
and the flux becomes continuous in the decompactification limit $\widetilde 
R\rightarrow\infty$.  

In our expression \eeintgsum\ for the thermodynamic potential, we can now 
simultaneously perform T-duality on all the Type 0A string theories 
parametrized by the value of $q$.  $\Gamma(R,\mu)$ is thus mapped to 
\eqn\eeintgsumd{\eqalign{2\pi R\cdot\Gamma
&=-\frac{1}{2}\sum_{q\in\Z}\Re\int\frac{d\sigma}{\sigma}\,e^{i\tilde\mu\sigma}
\frac{1}{\sinh(\sigma/2)}\frac{1}{\sinh[\sigma/(2\widetilde R)]}
e^{-|q|\sigma/(2\tilde R)}\cr
&=-\frac{1}{4}\Re\int\frac{d\sigma}{\sigma}\,e^{i\tilde\mu\sigma}
\frac{1}{\sinh(\sigma/2)}\frac{1}{\sinh^2[\sigma/(4\widetilde R)]}.\cr}}
In this way, the thermodynamic potential has been rewritten as an infinite 
sum over $q$ of contributions from {\it Type 0B vacua\/} with RR flux equal 
to $\nu=iq/(2\widetilde R)$.  Indeed, with the help of \newhat\ and 
\malsei , the individual terms in the sum in \eeintgsumd\ can indeed be 
recognized as the partition functions of the corresponding Type 0B vacua with 
flux $\nu$.  

As we pointed out in Section~2.1, the Hamiltonian of noncritical M-theory 
can also be written as an integral over a one-parameter family of Type 0B 
theories parametrized by the continuous eigenvalue $\nu$ of the 
single-particle Hamiltonian $h_2$ of (say) the second of the two 
one-dimensional oscillators.  When we put the theory on a Euclidean 
time circle of radius $\widetilde R$, the eigenvalues $\nu$ become 
purely imaginary and quantized in the units of $1/\widetilde R$, and the 
integral over $\nu$ turns into a discrete sum.  This suggests that 
the eigenvalue of $h_2$ in the Cartesian representation of the Fermi liquid 
describing noncritical M-theory should be identified with the Type 0B RR 
flux, much like the Type 0A RR flux $q$ was identified with the eigenvalue of 
the angular momentum $J$ in the polar-coordinate representation of the 
theory.  It also implies that noncritical M-theory inherits a T-duality 
symmmetry from the underlying T-duality between Type 0A and 0B strings.  
In noncritical M-theory, this T-duality swaps the Cartesian and 
polar-coordinate representations, in which the theory is decomposed into 
a collection of Type 0A or Type 0B string theories will all possible values 
of the corresponding RR flux.  

In the infinite temperature limit, $\tilde R$ goes to infinity, the Type 
0B flux $\nu$ becomes continuous, and the sum in \eeintgsumd\ turns into an 
integral.  In this limit, the density of states of the dual M-theory solution 
can be evaluated, and we obtain
\eqn\eedenslim{\widetilde\rho(\mudual)\equiv\frac{\p^2\widetilde\Gamma}{\p
\mudual^2}\sim-\frac{1}{2\pi}\log\cosh(\pi\mudual)+\frac{\Lambda}{2}.}
This reproduces our prediction \eedensihigh\ for the density of states of 
the effective high-temperature theory.  

Thus, our conclusions are as follows:

\smallskip
\item{$\bullet$}
Noncritical M-theory inherits a T-duality symmetry from the underlying 
T-duality between the Type 0A and 0B strings.  

\smallskip
\item{$\bullet$}
The high-temperature regime of the M-theory vacuum is described by the 
T-dual solution of noncritical M-theory, at the dual value of the 
temperature.  

Having established that the high-temperature regime is related to the 
low-temperature regime by T-duality, we can now return to the original task 
of Section~2.7 and study the high-temperature expansion of the thermodynamic 
potential $\Gamma$ of the original M-theory vacuum.  This is an expansion 
in the powers of $R$ with $\mudual$ fixed.  From the perspective of the 
T-dual theory, this is a low-temperature expansion in the powers of 
$1/\widetilde R$ with $\mudual$ fixed, analogous to the low-temperature 
expansion of the original vacuum discussed in Section~2.6.  We obtain
\eqn\eemudualrex{\eqalign{\frac{\p^3\Gamma}{\p\mudual^3}
&=-\frac{1}{8\pi R}\int_0^\infty d\sigma\,\sigma^2\,\sin(\mudual\sigma)\,
\frac{1}{\sinh^2(\sigma R/2)\,\sinh(\sigma/2)}\cr
&\approx \frac{1}{2\pi R^3}\sum_{k=0}^\infty\frac{(2k-1)B_{2k}R^{2k}}{(2k)!}
\int_0^\infty d\sigma\,\sin(\mudual\sigma)\,\frac{\sigma^{2k}}{\sinh(\sigma/2)}
\cr
&\approx \frac{1}{2\pi R^3}\sum_{k=0}^\infty
\frac{(2k-1)B_{2k}R^{2k}}{(2k)!}(-1)^k\,\frac{\p^{2k}}{\p\mudual^{2k}}
\int_0^\infty d\sigma\,\sin(\mudual\sigma)\,\frac{1}{\sinh(\sigma/2)}\cr
&\approx \frac{1}{2R}\sum_{k=0}^\infty
\frac{(2k-1)B_{2k}R^{2k-2}}{(2k)!}(-1)^k\,\frac{\p^{2k}}{\p\mudual^{2k}}
\,\tanh(\pi\mudual).\cr}}
This can be summarized in a compact formula analogous to our low-temperature 
expansion formula \eecompex , 
\eqn\eecompactd{\frac{\p^3\Gamma}{\p\mudual^3}\approx-\frac{\left(
\frac{\p}{\p\tilde\mu}\right)^2}{8R\,\sin^2\left(\frac{R}{2}\frac{\p}{
\p\tilde\mu}\right)}\,\tanh(\pi\mudual).}
The astute reader may recognize in \eemudualrex\ the strong resemblance to the 
partition function of the A-model topological closed string theory on the 
resolved conifold (see, \eg , \refs{\toporevmm,\toporevnv} for reviews), 
with $R$ essentially playing the role of the string coupling of the A-model, 
and $\mudual$ related to the K\"ahler modulus of the conifold.  This is our 
first hint of a much deeper relation between noncritical M-theory compactified 
on a Euclidean time circle, and the strings of the topological A-model.  This 
relation will be the topic of Section~3 below.  

\subsec{Noncritical M-Theory at the Debye Temperature} 

It was noticed in \ncm\ that the exact vacuum energy $E_0(\mu)$ of the vacuum 
solution in noncritical M-theory at zero temperature is mathematically 
equivalent to the partition function of Debye phonons in a Debye crystal at 
temperature $T_D=1/(2\pi)$,
\eqn\eevacdeb{E_0(\mu)=\frac{1}{2}\int_\mu^\Lambda d\xi\,\xi\,\rho_D(\xi)
\left(\frac{1}{e^{2\pi\mu}-1}+\frac{1}{2}\right),}
with the Debye density of states $\rho_D(\xi)\sim\xi$ characteristic of a 
$2+1$ dimensional system.  In a slight detour from the main theme of this 
paper, we shall now study the partition function of the noncritical M-theory 
vacuum at this ``Debye temperature.''  

The Debye temperature corresponds to the compactification radius $R=1$ of the 
Euclidean time dimension.  In noncritical string theory, this particular 
value of $R$ plays a special role, since the theory at this radius is related 
to the Penner model \penner .  

At $R=1$, the first derivative of the thermodynamic potential \eefreformal\ of 
our noncritical M-theory vacuum can be evaluated by residues, leading to 
\eqn\eedebyepart{\frac{\p\Gamma(\mu)}{\p\mu}=-\frac{1}{4}\left(\mu^2+
\frac{1}{4}\right)\tanh(\pi\mu)+\frac{\Lambda\mu}{2}+\frac{\Lambda^2}{2}.}
This can be further integrated, yielding the following exact expression for 
$\Gamma(\mu)$, 
\eqn\eedebinst{\eqalign{\Gamma(\mu)=&-\frac{1}{12}\,\mu^3+\frac{\Lambda}{4}
\,\mu^2+\left(\frac{\Lambda^2}{2}-\frac{1}{16}\right)\,\mu+C(\Lambda)\cr
&{}+\frac{1}{4}\sum_{k=1}^\infty (-1)^k\left(\frac{1}{k\pi}\,\mu^2
+\frac{1}{k^2\pi^2}\,\mu+\frac{1}{4k\pi}+\frac{1}{2k^3\pi^3}\right)
e^{-2k\pi\mu}.\cr}}
This formula nicely illustrates all of the features of the asymptotic 
expansion in the powers of $g_M$ that were discussed in Section~2.5 above.  
In particular, $\Gamma(\mu)$ contains a perturbative series which is one loop 
exact and starts off with the classical contribution at order $\mu^3=1/g_M^3$. 
This series is followed by an infinite series of instanton-like contributions. 
In each of the instanton terms, the perturbative quantum corrections are also 
one-loop exact, with the leading classical term vanishing, and the lowest 
nontrivial quantum correction starting at order $\mu^2=1/g_M^2$.    

Using \eedebyepart , and dropping the nonuniversal terms, $\Gamma(\mu)$ can 
also be rewritten in another interesting form, as 
\eqn\eedebyeferm{\Gamma(\mu)=\frac{1}{2}\int d\mu\,(\mu^2+\frac{1}{4})
\left(\frac{1}{e^{2\pi\mu}+1}-\frac{1}{2}\right).}
In this form, the universal part of the thermodynamic potential of noncritical 
M-theory at the Debye temperature is mathematically equivalent to the energy 
of a system of {\it fermions\/}	in $2+1$ dimensions at the Debye temperature, 
and with the effective density of states $\rho_D(\mu)\sim 2\mu+1/2\mu$.   
It is intriguing that this effective density of states is self-dual under an  
inversion of $\mu$.  

\newsec{Duality to the Closed String Theory of the Topological A-Model}

Our analysis of the partition function of noncritical M-theory at finite 
temperature in Section~2 has revealed a surprising connection with the 
amplitudes of the topological closed string of the A-model on the resolved 
conifold.  In this section, we will attempt to make this duality more precise, 
and analyze some of its possible implications.  

\subsec{First Comparison of the Partition Functions}

In Section~2.8, we derived the asymptotic high-temperature expansion of the 
partition function $\CZ_M$ of the noncritical M-theory vacuum, 
\eqn\eelogzsum{\frac{\p^3\log\CZ_M}{\p\mudual^3}\approx-\pi\sum_{k=0}^\infty
\frac{(2k-1)B_{2k}(iR)^{2k-2}}{(2k)!}\frac{\p^{2k}}{\p\mudual^{2k}}\left(
\tanh(\pi\mudual)\right).}
This can be rewritten in terms of polylogarithms as
\eqn\eelogzsumre{\frac{\p^3\log\CZ_M}{\p\mudual^3}\approx\frac{2\pi}{(iR)^2}
\left(\frac{1}{2}+\Li_0(-e^{-2\pi\tilde\mu})\right)-\sum_{k=1}^\infty
(2\pi iR)^{2k-2}\frac{(2\pi)^3B_{2k}}{2k(2k-2)!}\Li_{-2k}(-e^{-2\pi\tilde\mu})
.}
We now wish to integrate this expression to get an expansion of $\log\CZ_M$.  
\eelogzsumre\ determines $\log\CZ_M$ up to a second-order polynomial in 
$\mudual$, whose coefficients could be functions of $R$ containing 
non-universal dependence on the cutoff $\Lambda$.  The universal dependence 
of this polynomial on $R$ can be determined by taking derivatives of 
$\log\CZ_M$ with respect to $\beta$, in exact parallel with our analysis of 
the $R$ dependence of the polynomial \eepolynund\ in Section~2.3.  In the end, 
the nonuniversal cutoff dependence afflicts only a few lowest-order terms in 
the double expansion in $R$ and $\mudual$, and we obtain 
\eqn\eefinalz{\eqalign{\log\CZ_M\approx\frac{1}{(2\pi iR)^2}&\left(p(\mudual,
\Lambda)+\frac{(2\pi \mudual)^3}{12}-\Li_3(-e^{-2\pi\tilde\mu})\right)\cr
{}+&\left(C(\Lambda)-\frac{\pi\mudual}{12}-\frac{1}{12}\log(1+
e^{-2\pi\tilde\mu})\right)\cr
{}+&\sum_{k=2}^\infty(2\pi iR)^{2k-2}\frac{B_{2k}}{2k(2k-2)!}
\Li_{3-2k}(-e^{-2\pi\tilde\mu}).\cr}}
Here $p(\mudual,\Lambda)$ is a nonuniversal polynomial of second order in 
$\mudual$, and $C(\Lambda)$ is a nonuniversal constant; both $p(\mudual,
\Lambda)$ and $C(\Lambda)$ are independent of $R$.    

In this form, the close similarity to the partition function of the closed 
topological string of the A-model on the resolved conifold is apparent.  
Indeed, recall (or see in \refs{\gova,\govam,\toporevmm,\toporevnv}) that 
the partition function of the A-model on the resolved conifold can be written 
as an asymptotic expansion in the powers of the string coupling constant 
$g_A$, with coefficients being functions of the K\"ahler modulus $t_A$, 
\eqn\eelogza{\eqalign{\log\CZ_A&\approx\frac{1}{g_A^2}\left(p_A(t_A)+
\frac{t_A^3}{12}-\Li_3(e^{-t_A})\right)
+\left(C_A-\frac{t_A}{24}-\frac{1}{12}\log(1-e^{-t_A})\right)\cr
&{}+\sum_{h=2}^\infty g_A^{2h-2}\left(\frac{B_{2k}B_{2k-2}}{2k(2k-2)(2k-2)!}
+\frac{B_{2k}}{2k(2k-2)!}\Li_{3-2k}(e^{-t_A})\right).\cr}}
Here $p_A(t_A)$ is a non-universal second-order polynomial in $t_A$, and 
$C_A$ is a non-universal constant.  The real part of the K\"ahler modulus 
$t_A$ measures the size of the $S^2$ of the resolved conifold, and its 
imaginary part is given by the $B$-field flux through this $S^2$.  The 
$t_A$-independent terms in \eelogza\ come from constant maps from the 
string worldsheet to the target conifold, and the polylogarithms come from 
worldsheet instantons.  

A first comparison of \eefinalz\ and \eelogza\ suggests the identification
\eqn\eedualprel{g_A=2\pi i R,\qquad t_A=2\pi\mudual+\pi i.}
The matching between $\CZ_M$ and $\CZ_A$ is close, but there is an apparent 
difference: The constant term in \eelogza\ at each order in the A-model string 
coupling is absent in the partition function of the vacuum of noncritical 
M-theory.  

We claim that this difference is purely due to the fact that the two partition 
functions $\CZ_A$ and $\CZ_M$ are normalized differently.  Indeed, notice that 
the partition function of the A-model comes out naturally normalized so that 
the genus-$h$ term in $\CZ_A$ diverges as $t_A^{2-2h}$, all subleading terms 
going to zero as $t_A\rightarrow 0$.  This is in accord with the 
Gopakumar-Vafa duality \refs{\gova,\govam}, which maps the A-model to $U(N_c)$ 
Chern-Simons gauge theory on $S^3$.  In order to check this behavior, we can 
expand the polylogarithms around $t_A=0$, using the asymptotic expansion 
formula 
\eqn\eenormza{\eqalign{\Li_{3-2k}(e^{-t_A})&\approx\Gamma(2k-2)t_A^{2-2k}
+\sum_{n=0}^\infty\frac{(-1)^n\zeta(3-2k-n)}{n!}t_A^n\cr
&\approx(2k-3)!t_A^{2-2k}-\frac{B_{2k-2}}{2k-2}+ \CO(t_A),\qquad k=2,3, 
\ldots.\cr}}
Thus, the polylogarithm terms in \eelogza , which come from the sum over 
worldsheet instantons of genus $k$, have a leading divergence 
$\sim t_A^{2-2k}$, followed by a subleading constant term as $t_A$ goes to 
zero.   From the point of view of the Gopakumar-Vafa duality, the leading 
divergence is the nonperturbative term (discussed for example in \oovafa) 
which is not captured in the 't~Hooft expansion of the dual Chern-Simons 
theory.  At each order in $g_A$, the term originating from the constant maps 
in \eelogza\ is precisely such that it subtracts the subleading constant from 
the instanton polylogarithms, ensuring that -- order by order in $g_A$ -- 
$\log\CZ_A$ vanishes at $t_A=0$ after the leading $\sim t_A^{2-2h}$ divergence 
has been subtracted.  

On the other hand, the partition function of noncritical M-theory is naturally 
normalized so that it simplifies in a different limit, of 
$\mudual\rightarrow\infty$.  If it were not for the divergence in 
\eefreformal , setting $\mu=\infty$ in this expression would lead to 
$\CZ_M=1$.  This is again intuitively clear, since sending $\mu$ to infinity 
corresponds to emptying the Fermi sea; in the absence of any fermions, the 
partition function should be equal to one.  The divergence of $\log\CZ_M$ 
makes the precise realization of this formal expectation slightly subtle: 
One must empty the Fermi sea simultaneously with the double-scaling limit 
(\ie , to set $\mu\sim\Lambda$ as $\Lambda\rightarrow\infty$) in order to 
ensure that $\CZ_M=1$ as $\mudual\rightarrow\infty$.  

When we normalize $\CZ_M$ and $\CZ_A$ so that they are equal to one for the 
same (but otherwise arbitrarily chosen) fixed value $t_{A,{\rm fix}}$ of 
$t_A$, they turn out to be equal as asymptotic series in $t_A$.  In this 
sense, $\CZ_A$ and $\CZ_M$ carry precisely the same information, and we can 
summarize our result in the following duality relation between the partition 
functions:
\eqn\eedualzazm{\frac{\CZ_A(g_A,t_A)}{\CZ_A(g_A,t_{A,{\rm fix}})}
=\frac{\CZ_M(R,\mudual)}{\CZ_M(R,\mudual_{\rm fix})},}
with the parameters on the two sides related by \eedualprel\  (in particular, 
$t_{A,{\rm fix}}=2\pi\mudual_{\rm fix}+\pi i$).  

It is worth stressing again that since the A-model is only known as an 
asymptotic series in $g_A$, the duality relation \eedualzazm\ is to be only 
interpreted as an equality between two asymptotic expansions.  However, within 
the framework of noncritical M-theory, $\CZ_M$ is defined nonperturbatively in 
$R$, and therefore represents a possible nonperturbative completion of the 
topological A-model.  

\subsec{A More General Form of the Duality}

In the thermodynamic ensemble, the partition function corresponds to fermions 
antiperiodic around the Euclidean time dimension $t_E$.  
Since the thermodynamic interpretation is no longer important in this section, 
we shall relax this condition, and allow for an arbitrary phase $b$ in the 
periodicity of the fermions, 
\eqn\eepsiper{\Psi(t_E+2\pi R,\lambda_1,\lambda_2)=e^{ib}\Psi(t_E,\lambda_1,
\lambda_2).}
The case of the thermal ensemble studied in Section~2 corresponds to $b=\pi$.  

How does the general phase appear in the partition function?  
In the first-quantized framework that we used to calculate $\log\CZ$ in 
\eefreformal , allowing for the general phase in \eepsiper\ amounts to the 
insertion of an extra factor of $(-e^{ib})^F$, with $F$ the fermion number. 
(The minus sign is due to the fact that \eefreformal\ calculates $\log\CZ$ 
for the thermal case of $b=\pi$.)  Hence, the partition function with 
arbitrary periodicity of the fermions becomes 
\eqn\eefreformal{\log\CZ_M(R,\mudual,b)=\int_{-\infty}^{\infty}d\xi\,\rho(\xi)
\log\left(1-e^{ib}e^{2\pi\tilde\mu-2\pi R\,\xi}\right).}
Expanding this in the asymptotic series in $R$, we obtain
\eqn\eelogzb{\eqalign{\log\CZ_M\approx\frac{1}{(2\pi iR)^2}&\left(p(\mudual,b,
\Lambda)+\frac{(2\pi\mudual+ib)^3}{12}-\Li_3(e^{-2\pi\tilde\mu-ib})\right)\cr
+&\left(C(b,\Lambda)+\frac{2\pi\mudual+ib}{24}+\frac{1}{12}\log(1-
e^{-2\pi\tilde\mu-ib})\right)\cr
-\sum_{k=2}^\infty&(2\pi iR)^{2k-2}\frac{B_{2k}}{2k(2k-2)!}
\Li_{3-2k}(e^{-2\pi\tilde\mu-ib}).\cr}}
This again matches $\log\CZ_A$ modulo normalization, but now the entire range 
of complex values of the K\"ahler modulus $t_A$ is allowed, with the phase $b$ 
playing the role of the imaginary part of $t_A$.  

Recalling that in the original variables $\mudual$ is equal to $R\mu$, 
and that $g_M=1/\mu$ plays the role of the coupling constant in the 
noncritical M-theory vacuum, we find the duality relations
\eqn\eeduamarea{g_A=2\pi iR,\qquad t_A=2\pi\mudual+ib=\frac{2\pi R}{g_M}+ib}
advertized in the introduction.  With this matching of the parameters, the 
partition function are related by a generalization of \eedualzazm\ to the 
case of arbitrary $b$, 
\eqn\eedualzazmb{\frac{\CZ_A(g_A,t_A)}{\CZ_A(g_A,t_{A,{\rm fix}})}=
\frac{\CZ_M(R,\mudual,b)}{\CZ_M(R,\mudual_{\rm fix},b_{\rm fix})}.}
We have again chosen an arbitrary normalization point $t_{A,{\rm fix}}$, 
with $\mudual_{\rm fix}$ and $b_{\rm fix}$ related to $t_{A,{\rm fix}}$ via 
\eeduamarea .  

\subsec{Solutions with the Universal Bottom of the Fermi Sea}

As we have just seen, the partition functions of the vacuum of noncritical 
M-theory and of the topological A-model on the resolved conifold are virtually 
identical, differing only in their normalization properties.  Here we wish to 
point out the existence of a larger family of solution of noncritical 
M-theory, whose partition functions are naturally normalized in a way similar 
to that of the partition function of the A-model.  

The conventional vacuum of noncritical M-theory corresponds to a large $N$ 
limit of the system of $N$ fermions occupying all single-particle states from 
the cutoff $-\Lambda$ up to the top of the Fermi sea at the double-scaled 
Fermi energy $\mu$.  Recall that in Section~7.4 of \ncm , we studied a broader 
class of noncritical M-theory solutions, characterized by two Fermi surfaces: 
The universal top of the sea at some $\mu=\mu_+$, and the universal bottom 
of the sea at some $\mu_->\mu_+$ (see Figure~3).  This construction gives 
a two-parameter family of static solutions of noncritical M-theory, 
parametrized by $\mu_\pm$.%  
\fig{The solution of noncritical M-theory with a universal bottom of the 
Fermi sea.  In this solution, all single-particle states are occupied between 
the top top of the Fermi sea at $\mu_+$ and the bottom of the sea at 
$\mu_-$.  (As in Fig.~1, the angular dimension of the spatial plane has been  
suppressed.)}{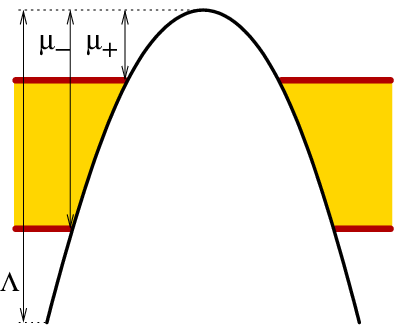}{2truein}

Consider now this family of solution in Euclidean signature, with the 
Euclidean time compactified on a circle of radius $R$, and define 
$\mudual_\pm=2\pi R\,\mu_\pm$.  The special case of particular interest to 
us is 
\eqn\eeset{\mudual_+=-\mudual,\qquad\mudual_-=0.}
Thus, the Fermi sea is filled from the top of the potential at zero, to some 
Fermi energy $\mu$ {\it above\/} the top of the potential.  

The partition function $\hat\CZ_M(R,\mudual)$ of this solution of noncritical 
M-theory no longer approaches one at $\mudual\rightarrow\infty$.  Instead, it 
simplifies in the limit of $\mudual\rightarrow 0$, which corresponds 
to emptying the Fermi sea.  It would be interesting to study this extended 
class of solutions of noncritical M-theory for general values of 
$\mudual_\pm$, and identify their relation to the topological A-model.  This 
is, however, beyond the scope of this paper.  

\newsec{Consequences of the Duality}

The proposed duality between noncritical M-theory and closed string theory 
of the topological A-model exhibits several notable features, and can be 
expected to have interesting implications for a wide range of phenomena 
traditionally related to the topological A-model.  Here we present some 
comments and raise some questions about the possible implications of this 
duality.  

As we saw in Section~3.2, the parameters on the two sides of the duality are 
related as follows, 
\eqn\eedrll{g_A=2\pi iR,\qquad t_A=\frac{2\pi R}{g_M}+ib.}
In particular, the duality relates the coupling constant $g_M$ of noncritical 
M-theory to the K\"ahler modulus of the resolved conifold on the dual side. 
Thus, the dimensionless coupling constant of noncritical M-theory acquires 
a geometric interpretation, as promised in Section~2.5.  

From the point of view of the dual A-model, the M-theory coupling constant 
$g_M$ plays the role of a worldsheet coupling.  In \ncm\ (and again in 
Section~2.5), we found the characteristic leading behavior $\sim 1/g_M^3$ of 
the free energy of the noncritical M-theory vacuum at weak coupling.  Our 
duality \eedrll\ provides an intriguing explanation for this leading power of 
$1/g_M^3\sim N^3$ in the large-$N$ behavior of noncritical M-theory, by 
relating it to the triple intersection form of the Calabi-Yau, which gives 
a tree-level contribution in the dual A-model \refs{\toporevmm,\toporevnv}.  

\medskip
{\it Topological M-theory for the A-model:}
\smallskip

The duality \eedrll\ relates the string coupling constant $g_A$ of the A-model 
to a geometric parameter -- the radius of the Euclidean time dimension -- on 
the dual side of noncritical M-theory.  Hence, the A-model string coupling 
$g_A$ acquires a geometric interpretation, as the size of the 
``M-theory circle.''  Such a relation $g_A\sim R$ between the string coupling 
and the radius of the M-theory circle is one of the hallmarks of the string 
theory/M-theory duality in critical string theory.  This suggests that 
noncritical M-theory could be a realization of the idea of ``topological 
M-theory,'' as proposed for closed topological string theory of the A-model 
on Calabi-Yau in \topomth\ (see also \topozth).  

In this context, it is intriguing that the role of the M-theory dimension is 
played by the Euclidean {\it time\/} dimension of noncritical M-theory.  One 
motivation for introducing topological M-theory was to explain the apparent 
behavior of the partition function of the A-model as a wave function of a 
system evolving in an extra time dimension.  Perhaps, the factor of $i$ in the 
relation \eedrll\ between the string coupling $g_A$ and the radius $R$ is an 
indication that the duality should be more naturally interpreted after the 
Wick rotation to real time, effectively replacing $R$ with $iR$ and 
reinterpreting $R$ as the time lapse in real time evolution.  

\medskip
{\it Relation to Chern-Simons theory on $S^3$:}
\smallskip

Gopakumar-Vafa duality \refs{\gova,\govam} relates the A-model on the resolved 
conifold to the $U(N_c)$ Chern-Simons gauge theory on $S^3$.  The 
Gopakumar-Vafa relations are
\eqn\eegvdual{g_A=\frac{2\pi i}{k+N_c},\qquad t_A=\frac{2\pi i N_c}{k+N_c},}
where $k$ is the level of the Chern-Simons theory.  

When combined with the Gopakumar-Vafa duality, our duality implies a relation 
between noncritical M-theory and $U(N_c)$ Chern-Simons gauge theory on $S^3$.  
For the quantities in noncritical M-theory, this implies
\eqn\eegvmdual{R=\frac{1}{k+N_c},\qquad \mu+\frac{ib}{2\pi R}={iN_c}.}
These relations have a very natural interpretation, with $R$ being the 
parameter of the large-$N_c$ expansion, \ie , the gauge coupling constant 
$g_{CS}^2$.  Moreover, if we choose to perform the Wick rotation to real 
time, \eegvmdual\ implies an intriguing relation $g_M=1/N_c$, identifying the 
M-theory coupling constant with the inverse number of colors in the dual 
Chern-Simons gauge theory.  Note also that since the Chern-Simons theory can 
be reinterpreted as a matrix model \csmm , this connection leads to a matrix 
model description of noncritical M-theory.  

\medskip
{\it Nonperturbative completion of the A-model:}
\smallskip

Since the partition function of the topological strings of the A-model is 
defined only as an asymptotic expansion in $g_A$, relations between partition 
functions implied by \eedualzazm\ are understood as equalities between 
asymptotic expansions.  However, the partition function $\CZ_M$ of noncritical 
M-theory is nonperturbatively defined, and therefore \eedualzazm\ can be 
interpreted as a nonpertutbative completion of the asymptotic series for the 
A-model.  

Another nonperturbative completion of the A-model has emerged recently in the 
context of the OSV conjecture \osv .  This conjecture relates the entropy 
of certain supersymmetric black holes to the absolute value squared of a 
topological string partition function, and its precise form is still being 
developed.  It should be interesting to investigate the relation between 
the nonperturbative completions of the A-model via the OSV conjecture and via 
the duality to noncritical M-theory.  In particular, the OSV conjecture has 
been tested in \refs{\vafaq,\aosv} for the topological string on a noncompact 
Calabi-Yau given by the $\CO(-p)\times\CO(p+2g-2)$ fibration over a genus-$g$ 
Riemann surface $\Sigma_g$.  The partition functions are related to the large 
$N_c$, $q$-deformed version of Yang-Mills on $\Sigma_g$.  In turn, 
two-dimensional large-$N_c$ Yang-Mills theory on $\Sigma_g$ is known to have 
an interpretation in terms of a string theory \grosst , whose worldsheet 
description is given by the topological rigid string \rigid\ (see also 
\cmr ).  The case of $p=1$ and $g=0$ is rather singular, and somewhat outside 
of the scope of \aosv .  In particular, the large-$N_c$ Yang-Mills theory on 
$S^2$ is known to exhibit 
the Douglas-Kazakov phase transition \doukaz .%
\foot{In the context of the OSV conjecture, this phase transition has been 
recently studied in \dkosv .}

However, $p=1$ and $g=0$ is the resolved conifold, for which the duality 
to noncritical M-theory represents a nonperturbative completion of the 
A-model, and could therefore be complementary to the methods of \aosv .  
Note in particular that the partition function \eefreformal\ of 
noncritical M-theory exhibits a singular behavior at $b=0$, with each term of 
the asymptotic expansion \eelogzb\ divergent at $2\pi\mudual+ib=0$.  We expect 
this singular behavior to be related to the Douglas-Kazakov phase transition 
of Yang-Mills theory on $S^2$.    

\medskip
{\it S-duality of the A-model from T-duality of noncritical M-theory:}
\smallskip
 
As we saw in Section~2, noncritical M-theory can be formally decomposed into 
an infinite number of sectors equivalent to Type 0A or 0B theory with all 
possible values of their RR flux.  When compactified on the thermal circle, 
noncritical M-theory inherits a duality symmetry from the T-duality 
between the Type 0A and 0B string theories.  This duality was used in 
Section~2.8 to shed light on the high-temperature behavior of noncritical 
M-theory. It acts on the parameters of the vacuum by
\eqn\eetduaag{\mudual=\mu\,R,\qquad \widetilde R=\frac{1}{2 R}.}

When combined with the duality \eedrll\ to the topological A-model, this 
T-duality predicts the existence of an S-duality for the A-model.  
Indeed, in terms of the A-model variables, \eetduaag\ acts by inverting 
the string coupling constant and rescaling the K\"ahler modulus, 
\eqn\eetduaga{g_A\rightarrow \frac{1}{g_A},\qquad t_A\rightarrow t_Ag_A,}
\ie , as an S-duality transformation.  

An S-duality for the A-model on a Calabi-Yau has been proposed in the 
literature \refs{\neitzvafa,\toposdual}.  It also acts by \eetduaga , and 
relates the A-model to the B-model on the same target and with its string 
coupling given by $g_B\sim 1/g_A$.  It should be interesting to investigate 
the relationship between the S-duality suggested by noncritical M-theory and 
the S-duality proposed in \refs{\neitzvafa,\toposdual}.  The latter has been 
related to T-duality in critical string theory in \skapustin .

\medskip
{\it The melting crystal and quantum foam:}
\smallskip

At infinite $t_A$, string theory of the topological A-model has been related 
to a classical crystal melting problem \foam .  The inverse temperature of 
that crystal is proportional to $g_A$.  In combination with our duality, 
\eedrll\ shows that the temperature of the melting crystal is essentially the 
temperature of the thermal ensemble in the noncritical M-theory vacuum.  
Moreover, in \ncm , we evaluated the exact vacuum energy of the noncritical 
M-theory vacuum, and found a surprising similarity with the partition function 
of phonons in a Debye crystal, with $\mu$ related to the size of this 
crystal.  \eedrll\ relates $\mu$ to the K\"ahler modulus of the conifold, 
which was set to infinity in \foam , corresponding to an infinite crystal in 
\foam .  The case of finite $t_A$ was studied in \okuda , with $\mu$ indeed 
providing a scale related to the size of the crystal.  Given this matching of 
the physical interpretation of the parameters, it is natural to suspect that 
the constituents of the spacetime foam crystal of \foam\ are intimately 
related to the constituent fermions of noncritical M-theory.  

\medskip
{\it Embedding into M-theory:}
\smallskip

The relation between noncritical M-theory and the topological A-model was 
found in Section~3 by comparing the partition functions of the two theories.  
It would clearly be desirable to have a more systematic derivation of such a 
duality, by embedding noncritical M-theory to superstring theory or 
eleven-dimensional M-theory.  

In the related case of the Gopakumar-Vafa duality, such a derivation exists 
\atmava .  It takes advantage of a relation between topological strings on 
Calabi-Yau and M-theory on a noncompact $G_2$ holonomy manifold.  From the 
eleven-dimensional vantage point of M-theory, the duality is a simple 
consequence of the flop transition.  In addition, $G_2$ holonomy manifolds are 
also central ingredients in the conjectured relation of topological strings on 
Calabi-Yau to topological M-theory in seven dimensions.  

We expect M-theory on $G_2$ holonomy manifolds to be relevant for the 
duality between noncritical M-theory and the topological A-model as well.    
Indeed, in the case of the Gopakumar-Vafa duality \atmava , the $G_2$ holonomy 
7-manifold in question is given by a quadratic equation in $\C^4$, 
\eqn\eegtwo{|z_1|^2+|z_2|^2-|z_3|^2-|z_4|^2=2\mu.}
This should be compared to the Fermi surface of the vacuum state in 
noncritical M-theory.  In the double scaling limit, the Fermi liquid 
describing noncritical M-theory in $2+1$ dimensions becomes semiclassical.   
The semiclassical Fermi surface is a hypersurface in the classical phase 
space parameterized by coordinates $\lambda_1,\lambda_2$ and momenta 
$p_1,p_2$.  In particular, the vacuum state of noncritical M-theory (at zero 
temperature) is described by the Fermi surface satisfying%
\foot{For other solutions of the classical equations of motion of the Fermi 
surface in noncritical M-theory, see \ncm .}
\eqn\eefsquad{p_1^2+p_2^2-\lambda_1^2-\lambda_2^2=2\mu.}
Thus, the Fermi surface of the noncritical M-theory vacuum corresponds to the 
intersection of the $G_2$ holonomy 7-manifold \eegtwo\ with the real section 
$\R^4\subset\C^4$!  This is very similar to the embedding of two-dimensional 
noncritical string theories into full string theory \refs{\ghvafa,\mina}.  
Interestingly, the flop transition of the $G_2$ holonomy manifold is realized 
by $\mu\rightarrow-\mu$, which in \eefsquad\ corresponds precisely to the 
particle-hole duality in noncritical M-theory.  

\newsec{Conclusions}

Noncritical M-theory in $2+1$ dimensions represents a unifying framework, 
in which various string theories are related by dualities, in a controlled 
setting of an exactly solvable model.  In this paper, we have seen further 
evidence supporting this picture.  In particular, we have found that 
noncritical M-theory is related to string theories in at least two different 
ways, via two distinct forms of M-theory/string theory duality.  

When compactified on the Euclidean time circle, noncritical M-theory in 
$2+1$ dimensions has two $U(1)$ symmetries: the rotations of the spatial 
eigenvalue plane, and the translations of the Euclidean time circle.  
The reduction on the angular $S^1$ on the eigenvalue plane is related to 
noncritical $\hat c=1$ Type 0A theory.  In this correspondence, the 
Kaluza-Klein charge of the reduction (\ie , the angular momentum on the 
eigenvalue plane) is identified with the RR charge of stable D0-branes in 
Type 0A string theory.  This was the original picture that led to the 
definition of noncritical M-theory in terms of a Fermi liquid in \ncm .  

In Section~3 of this paper, we have seen evidence that the reduction of 
noncritical M-theory along the Euclidean time circle is also related to a 
string theory, namely the closed string theory of the topological A-model on 
the resolved conifold.  In this correspondence, the radius $R$ of the 
Euclidean time $S^1$ is interpreted as the string coupling $g_A$ of the 
A-model.  

Among the most intriguing features of noncritical M-theory (and of its 
cousins, matrix models of noncritical string theories) is the fact that the 
physical spacetime emerges as a derived concept, associated to the existence 
and geometrical properties of the Fermi surface.  In this emergent picture 
of spacetime, the elementary fermions of the Fermi liquid -- originating from 
D0-branes in the underlying string theory -- are the fundamental constituents, 
and the smooth macroscopic effective geometry of spacetime is a collective 
phenomenon.  The rigid eigenvalue plane populated by the fermions is an 
auxiliary structure, only rather indirectly related to the physical space.  
Thus, the exactly solvable setting of noncritical M-theory seems particularly 
suitable for extracting more lessons about the emergence and microscopic 
constituents of spacetime in string and M-theory.  

\bigskip
\bigskip
\noindent{\bf Acknowledgements}
\medskip
We wish to thank Mina Aganagic, Ofer Aharony, Sumit Das, Eric Gimon and 
Aleksey Mints for useful discussions.  This material is based upon work 
supported by NSF grant PHY-0244900, DOE grant DE-AC02-05CH11231, an NSF 
Graduate Research Fellowship, and the Berkeley Center for Theoretical 
Physics.  

\listrefs
%%%%%%%%%%%%%%%%%%%%%%%%%%%%%%%%%%%%%%%%%%%%%%%%%%%%%%%%%%%%%%%%%%%%%%%%%%%%%
\end